# First year of coordinated science observations by Mars Express and ExoMars 2016 Trace Gas Orbiter

A. Cardesín-Moinelo[1], B. Geiger[1], G. Lacombe[2], B. Ristic[3], M. Costa[1], D. Titov[4], H. Svedhem[4], J. Marín-Yaseli[1], D. Merritt[1], P. Martin[1], M.A. López-Valverde[5], P. Wolkenberg[6], B. Gondet[7]
and Mars Express and ExoMars 2016 Science Ground Segment teams

*[1] European Space Astronomy Centre, Madrid, Spain*

*[2] Laboratoire Atmosphères, Milieux, Observations Spatiales, Guyancourt, France*

*[3] Royal Belgian Institute for Space Aeronomy, Brussels, Belgium*

*[4] European Space Research and Technology Centre, Noordwijk, The Netherlands*

*[5] Instituto de Astrofísica de Andalucía, Granada, Spain*

*[6] Istituto Nazionale Astrofisica, Roma, Italy*

*[7] Institut d'Astrophysique Spatiale, Orsay, Paris, France*

## Abstract

Two spacecraft launched and operated by the European Space Agency are currently performing observations in Mars orbit. For more than 15 years Mars Express has been conducting global surveys of the surface, the atmosphere and the plasma environment of the Red Planet. The Trace Gas Orbiter, the first element of the ExoMars programme, began its science phase in 2018 focusing on investigations of the atmospheric composition with unprecedented sensitivity as well as surface and subsurface studies. The coordination of observation programmes of both spacecraft aims at cross calibration of the instruments and exploitation of new opportunities provided by the presence of two spacecraft whose science operations are performed by two closely collaborating teams at the European Space Astronomy Centre (ESAC).

In this paper we describe the first combined observations executed by the Mars Express and Trace Gas Orbiter missions since the start of the TGO operational phase in April 2018 until June 2019. Also included is the science opportunity analysis that has been performed by the Science Operation Centres and instrument teams to identify the observation opportunities until the end of 2020. These results provide a valuable contribution to the scientific community by enabling collaborations within the instrument teams and enhance the scientific outcome of both missions. This information is also valuable to other Mars missions (MAVEN, Mars Reconnaissance Orbiter, Curiosity, …), which may be interested in observing these locations for wider scientific collaboration.

In this work we have analysed the simultaneous and quasi-simultaneous opportunities for cross-calibrations and combined observations by both missions, in particular for the vertical atmospheric profiles with solar occultation and the global atmospheric monitoring with nadir observations. As a result of this work we have identified simultaneous solar occultations that can be combined to compare vertical atmospheric profiles of the same region observed by different instruments within less than 2 minute difference, and many other quasi-simultaneous occultation observations within 15 minutes difference at various latitudes and local times. We have also analysed and identified the simultaneous nadir observations that have been planned regularly at different distances and illumination conditions, and studied the seasonal evolution of the orbital crossing points and the alignment of the ground tracks driven by seasonal orbital variations.

Lastly we provide an analysis of future opportunities to improve the global coverage of the atmosphere until the end of 2020. This study includes combined opportunities for nadir, solar occultations and a preliminary study of the coverage for limbs and radio occultations between both spacecraft. The resulting observations strongly increase the robustness of both Mars Express and Trace Gas Orbiter investigations due to cross-calibration of the instruments. They significantly increase spatial and temporal coverage, open new opportunities for scientific collaboration and synergy thus enhancing overall science return of both missions.

## 1. Introduction

After 15 years in orbit, the Mars Express (MEX) orbiter is still fully operational. It has been providing a wealth of data since its insertion into polar elliptical orbit around Mars in December 2003. The Science Operations Centre at the European Space Astronomy Centre (ESAC), together with the instrument teams, is routinely working to maximize the scientific return of the mission. This covers a wide range of science objectives: surface and sub-surface geology, atmospheric structure, dynamics and composition, up to the plasma environment and the escape in the magnetosphere and the characterization of the Martian system including its two moons Phobos and Deimos. [Chicarro, 2004; Titov, 2017]

The ExoMars 2016 Trace Gas Orbiter (TGO) is the first mission of the ExoMars programme [Vago, 2015]. It was successfully inserted into a highly elliptic orbit around Mars in October 2016. This was followed by a long aerobraking phase that brought the satellite into its science orbit in March 2018. This milestone kicked off the start of the commissioning phase followed by the start of the nominal science phase in April 2018, when the first science operations took place after many years of development and planning. The science planning is coordinated by the TGO Science Operations Centre (SOC) at ESAC near Madrid, to implement all of the science observations and fulfil the scientific goals of the mission: atmospheric trace gases, climatology, surface geology and shallow subsurface water detection. [Metcalfe, 2018]

Mars science has a long history of collaboration between different international teams and missions. In the frame of the Mars Express mission we have seen how the European community has grown exceptionally over the years, reaching a prime role within the international community and establishing long lasting synergies. This has been very productive, as demonstrated by the various publications in this special issue and the EU H2020 project [UPWARDS]. We can see that the scientific collaborations between different Mars missions have been established over a wide range of topics, typically at the level of individual teams or research groups interacting and sharing their results directly within their specific area of investigation: subsurface (MEX/MARSIS and MRO/SHARAD), surface geology and mineralogy (MEX/OMEGA and MRO/CRISM, MEX/HRSC and MRO/HiRiSe), atmosphere (most cameras, spectrometers and radars) and up to the magnetosphere (MEX/ASPERA and MAVEN). Most of these scientific collaborations have been performed a-posteriori comparing the final results obtained by the different instruments with few efforts dedicated to coordinate specific observations a-priori. Some coordinated observations have been performed, including specific surface targets (for example the hydrated minerals [Carter, 2013]), special events like the passage of comet Siding Spring [GRL Special Issue, 2015], specific campaigns for Methane detection measurements [Giuranna, 2019a], dedicated support to the entry, descent and landing of several missions (Phoenix, Mars Exploration Rovers, Curiosity, Insight) and a few other coordinated observations with Earth-based telescopes [Aoki, 2015]. The success of these coordinated observations demonstrate the value of the collaboration between Mars Express and ExoMars TGO communities and the need for extra-efforts, at mission level, to enhance the coordination of joint observations.

In this paper we first give a short summary of each mission and its mission profile (sections 2 and 3), with the characteristics of each orbit and the differences that drive the observational capabilities. We then briefly describe how the opportunities identified in this study are used during the science planning (section 4), the synergistic capabilities between the instruments and the observations that can be combined to enhance the scientific outcome of both missions (section 5). The next sections contain the analysis of combined solar occultations (section 6), combined nadir observations (section 7), orbit alignment periods (section 8) and other future opportunities (section 9), followed by the final conclusions (section 10).

We provide details on the science opportunity analysis, using various operational tools inherited from previous planetary missions: SPICE, Cosmographia [Acton, 2018; Costa, 2018], SOLab [Costa, 2012] and MAPPS [van der Plas, 2016]. These various tools are used to perform geometrical and operational simulations of both spacecraft, taking into account the observation requirements of all instruments and the operational requirements for feasibility checks.

The resulting observations presented in this paper are relevant for the scientific collaborations and synergies within the teams working for both missions and the overall scientific community including other Mars missions. The observation times and locations listed in this paper offer excellent chances to learn and improve the knowledge not only about the performance of the instruments (through cross-calibrations) but also for inter-comparison of the interpretation of the data retrievals and the overall complementarity of the science returns.

## 2. The Mars Express Mission

Launched in June 2003, the Mars Express (MEX) spacecraft has been continuously returning large volumes of science data since its insertion in Mars orbit in December 2003. While the nominal mission was initially planned to last 687 Earth days (one full Martian year), it has now been operating continuously for over 15 Earth years. It has a capability to cover a very diverse set of science objectives using a well matched set of instruments: a High Resolution Stereo Camera (HRSC) mainly used for geomorphology and surface mapping [Jaumann, 2007]; a hyperspectral imaging spectrometer in the Visible and Infrared for surface mineralogy and atmospheric analysis (OMEGA) [Bibring, 2004]; two spectrometers devoted to atmospheric gas and aerosols in the infrared (Planetary Fourier Spectrometer – PFS and Ultraviolet and Infrared Atmospheric Spectrometer - SPICAM) [Formisano, 2005; Montmessin, 2017]; a wide angle context camera to monitor Martian climate and meteorology (Visual Monitoring Camera - VMC) [Ormston, 2011]; a Mars Sub-surface Sounding Radar Altimeter (MARSIS) [Picardi, 2010]; a magnetospheric plasma instrument (ASPERA) [Barabash, 2006] and finally a Radio Science investigation (MaRS) [Pätzold, 2016] that uses the communication radio link to sound the neutral atmosphere and ionosphere, the surface and solar corona.

The longevity and richness of the scientific output of Mars Express is to a large extent a result of its orbital characteristics and the high flexibility of the spacecraft to observe under a large variety of conditions. The Mars Express spacecraft is in a highly elliptical polar orbit, with an inclination of 86° and a period of nearly 7 hours, resulting in about three orbits per day. The pericentre height is approximately 350 km, while the apocentre height is approximately 10,000 km. The orbit is not synchronized with Mars, Earth or the Sun, and so it is drifting according to celestial mechanics with a slow precession of the pericenter latitude and the illumination conditions. The evolution of the pericentre is shown in Figure 1, with plots of the sub-spacecraft latitude and solar elevation angle, defined as the angle between the horizon line and the sun (0 degrees at terminator, positive angles are illuminated up to 90 degrees at the sub-solar point, negative values are in the dark with a minimum of -90 degrees in the anti-solar point). The variation in latitude has a period of nearly 20 months, while the solar elevation angle evolution has a period of 8~12 months. These define long science observation campaigns for the various regions of the planet based on scientific objectives.

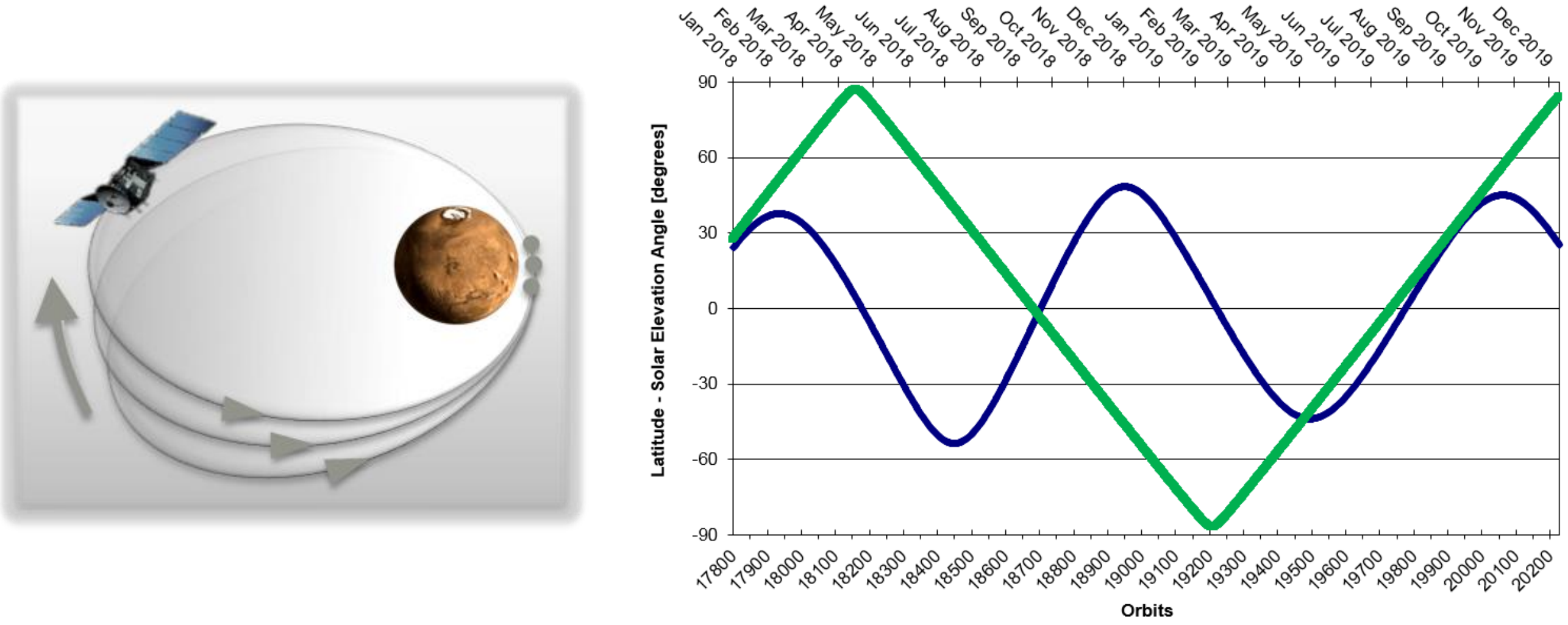

**Figure 1 Illustration of the Mars Express orbit precession around Mars (left), long term evolution of the latitude of the pericentre (green) and solar elevation angle at the pericentre (blue) of each orbit over 2018-2019 (right).**

The high eccentricity of the orbit provides a very wide range of distances between the spacecraft and Mars, that allows for the observation of the planet with varying observing conditions. However, the long-term evolution of the orbit precession ensures very stable and slow changing seasons, with very slow latitudinal variations and illumination at pericentre during which long observation campaigns (3~6 months) can be planned.

In terms of operations, the Mars Express spacecraft has a wide range of technical rules and constraints that must be respected during the science planning process to ensure the safety of the mission. These operational constraints cover all subsystems of the spacecraft: electrical power (solar array orientation towards the sun, discharge of the batteries in eclipse, etc), communications (ground station availability, variable downlink bitrate, instrument data rates, memory management, etc), thermal (accumulated heat flux on the spacecraft panels, illumination of various sensors, etc), propulsion (manoeuvres and maintenance wheel off-loadings), commanding limitations, etc.

Despite the high number of constraints, the good performance of the spacecraft and the gained experience by the operations teams throughout the years has allowed a great flexibility to point almost anywhere in space and operate almost at any time along the orbit. The main limitation is the high-gain antenna, which is fixed with respect to the spacecraft, meaning that the communication passes with ground stations have to be performed in Earth pointing mode. Therefore, the science instruments cannot point to Mars all the time, and the science pointing blocks have to be accommodated appropriately outside the communication passes. In general, the science timeline has typically 2~3 science pointing blocks per orbit, with a total time of ~1.5-3 hours per orbit dedicated to remote sensing observations, while the rest is devoted to Earth communications. This corresponds to an approximate science operations duty cycle of 20-40%, with high variability depending on the season. Note that this does not apply to the ASPERA instrument as it is operated continuously except for spacecraft maintenance slots or limited resource periods, and the Radio Science experiment that is operated mostly during the Earth communication passes.

## 3. The ExoMars 2016 Trace Gas Orbiter Mission

ExoMars 2016 Trace Gas Orbiter is the first mission of the ExoMars Programme, developed jointly by the European Space Agency (ESA) and Roscosmos (Russia). The main scientific goal of this mission is to monitor the atmospheric minor species that are present in small concentrations (less than 1% of the atmosphere) but nevertheless are key to the understanding the state of the Martian atmosphere and its evolution. Methane is of particular interest as it could be an evidence for possible biological or geological activity. This investigation is supported by high-resolution imaging of the surface, in particular to find possible sources of trace gases, and the mapping of the water content in the shallow the sub-surface.

The Trace Gas Orbiter (TGO) carries a scientific payload capable of addressing the mission's scientific goals. For the detection and characterisation of trace gases in the Martian atmosphere there are two sets of spectrometers: the Atmospheric Chemistry Suite (ACS) [Korablev, 2018] and the Nadir and Occultation for MArs Discovery (NOMAD) [Vandaele, 2018]. Both sound the Martian atmosphere in solar occultation, nadir and limb modes in order to obtain high-resolution vertical profiles of various gases and aerosols, mapping the atmospheric conditions of the whole planet with an unprecedented accuracy and sensitivity compared to previous missions. The Colour and Stereo Surface Imaging System (CASSIS) obtains high-resolution colour and stereo images of selected targets on the surface [Thomas, 2018]. Finally the Fine Resolution Epithermal Neutron Detector (FREND) [Mitrofanov, 2018] maps the subsurface hydrogen to a depth of one meter, to reveal any deposits of water-ice hidden just below the surface which, along with locations identified as sources of the trace gases, and stereo colour imaging, could influence the choice of landing sites of future missions.

The nominal science orbit of TGO around Mars is nearly circular with a two-hour period, at approximately 400 km distance from the surface, with an inclination of 74° and a characteristic node regression that makes the orbit plane rotate around the planet with a typical cycle of approximately 7 weeks. The evolution of the beta angle (the angle between the orbit plane and the Sun), is shown in Figure 2 from [Geiger, 2018]. This beta angle drives the main seasons of the mission and defines the long-term planning campaigns for solar occultations and nadir observations. We note in particular the high beta angle seasons (>68 degrees), in which no solar occultations occur, and the low beta angle seasons (<31 degrees), which are the best for solar occultations but require spacecraft flips that may potentially affect the pointing capability.

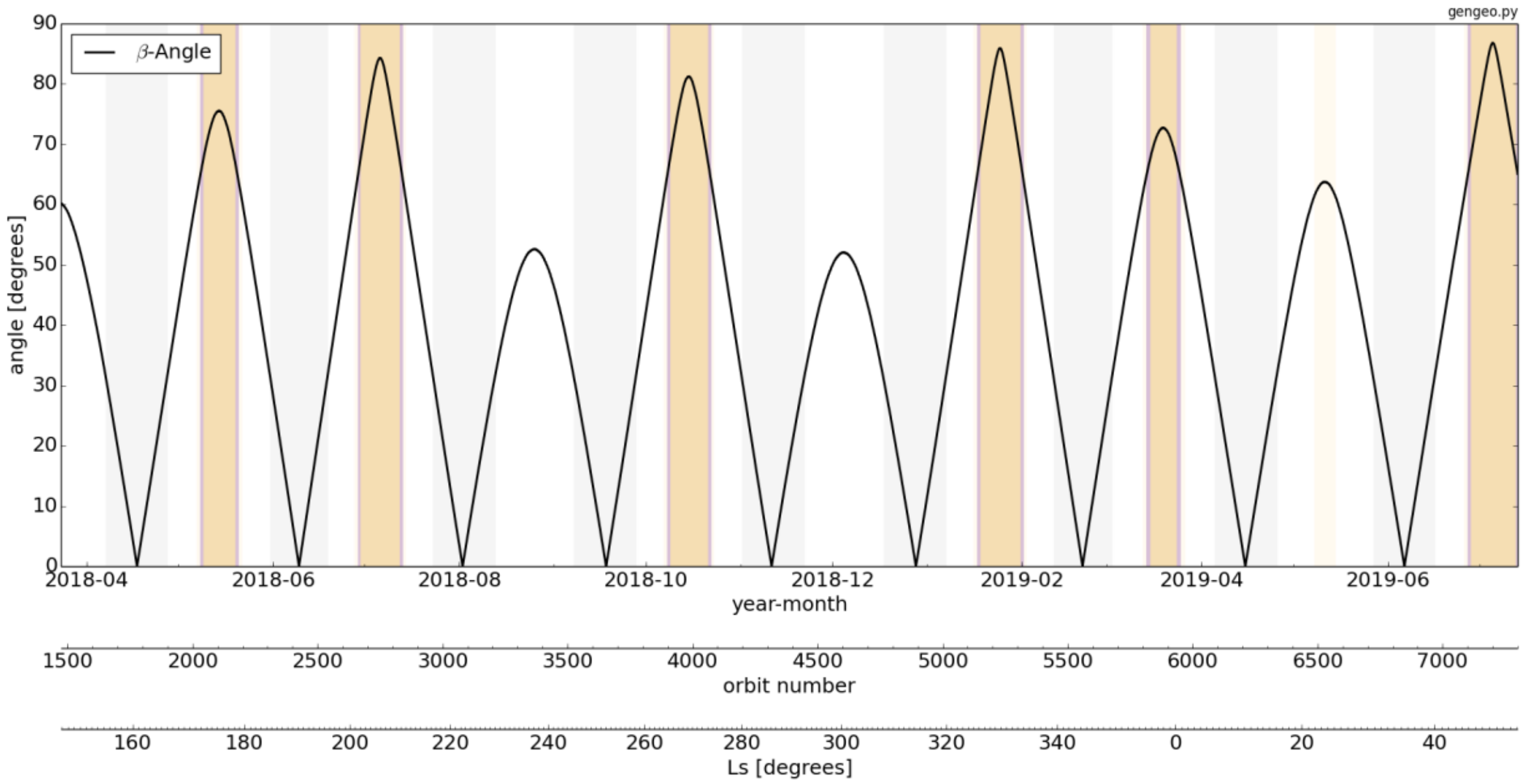

**Figure 2**. **Long-term evolution of the TGO orbit Beta Angle (angle between the orbital plane and the Sun) with respect to date, orbit number and Mars Solar Longitude (Ls). The brown-shaded regions specify high beta angle periods (>68 degrees) where no solar occultations occur. The grey shaded regions specify the periods with beta angle (<31 degrees) for which rate-controlled spacecraft flips are required.**

A major advantage of the ExoMars TGO spacecraft compared to Mars Express is the capability to position the high-gain antenna to track the Earth while the spacecraft is pointing to Mars. This allows for almost continuous science observations, assuming nadir pointing by default and otherwise pointing the solar occultation channels to the Sun whenever needed for solar occultation measurements. There are however several limitations due to the satellite maintenance slots, rate-controlled spacecraft flips to maintain the solar power optimization (with beta angle <31deg as shown in Figure 2), and numerous other constraints defined by the Mission Operations Centre. However, the science operations duty cycle is still very high compared to Mars Express (>75%). [Geiger, 2018]

In general, the TGO orbit long-term evolution is very dynamic compared to Mars Express, and has observing seasons that vary regularly on a weekly basis, due to the orbital node regression. This allows for a full surface and local time coverage on a monthly basis, except for the polar regions.

We note here that TGO also has an important role as a data relay for landers and rovers on the surface, currently supporting the NASA rovers and in the future also the ExoMars 2020 Rover and Surface Platform. We can anticipate that the science operations of TGO are expected to continue for many years, although the impact of the relay capability in the TGO science plan after 2021 is still under discussion and will only be confirmed as part of the mission extension phase, currently in preparation.

## 4. Planning of combined science opportunities and observation execution

We will now briefly describe the main planning cycles coordinated by the science operations centres at ESAC and the work within these centres. This includes the science opportunity analysis given in this paper until the final confirmation of instrument commands that are delivered to the spacecraft to execute the coordinated science observations.

As soon as the reference spacecraft trajectories are provided by Flight Dynamics to the Science Operations Centres the combined opportunities described in this paper are computed as part of the Long Term Planning cycle approximately 6 months in advance of the execution period [Geiger, 2018; Cardesin, 2018]. This is followed by geometry analysis and discussion within the science teams, defining the long-term strategy for the upcoming mission phase. We note that the typical accuracy of the long-term trajectory is approximately 10 seconds (phasing error of the spacecraft position within the orbit), which is sufficient for the long-term planning campaigns and also for the joint MEX-TGO analysis described in this paper.

The confirmation of instrument observations is performed during the Medium Term Planning (MTP) cycle, which encompasses a 28-day period of observations, and is planned approximately 3 months in advance of the execution [Ashman, 2018]. The joint opportunities identified at the LTP stage are used here as a starting point for the instrument teams to select their top science priorities, which are then discussed, iterated among all the teams and harmonized to fulfil the scientific requirements while respecting the operational constraints of the mission. Subsequently, the observation plan is verified by the Mission Operations Centre (MOC) at the European Space Operations Centre in Germany, and the pointing and commanding of all instruments is confirmed.

Finally, the Short Term Planning (STP) cycle is used to confirm the detailed commanding of all payloads on-board the respective satellites [Ashman, 2018]. This iteration is done on a weekly basis, and confirmed just 10 days prior to the uplink of the payload commands to the spacecraft by the MOC. In this cycle the satellite pointing remains unchanged, only the instrument parameter settings are optimized for observing different gas species or changing the integration times depending on scientific objectives and the latest information on geometrical conditions.

## 5. Coordinated science observations

### 5.1. Scientific rationale

As previously mentioned, the motivation behind this study is to promote the scientific collaborations and synergies between the teams working for Mars Express, Trace Gas Orbiter and the overall scientific community.

In this paper, we discuss the instruments and observation opportunities that can be combined to enhance the scientific outcome of the instruments. In particular, we focus here on the remote sensing spectrometers, whose data can be combined to learn more about the Martian atmospheric circulation. Table 1 provides a brief summary of the atmospheric parameters that can be measured by the different spectrometers on-board Mars Express and TGO, including atmospheric temperatures, dust, $CO_2$ ice, water ice, water vapour, CO and methane. Note this table is only meant as an indicative summary of the possible synergies and it is not meant to be an exhaustive list of all atmospheric parameters or minor species, which would be out of the scope of this work.

The coordinated measurements discussed in this paper can be used to enhance significantly the spatial and temporal coverage of the atmospheric investigations, and allow for cross-calibration of the data to improve the retrievals and interpretation of the results. The collaboration between various instruments for atmospheric studies has already been demonstrated multiple times in the past. As an example, the oxygen nightglow at 1.27μm was first detected on Mars by OMEGA [Bertaux, 2011], then studied further by SPICAM [Fedorova 2012] and will now be extended with NOMAD and ACS to achieve higher vertical resolution and extended spatial coverage. More recently, [Montmessin, 2018] applied a dedicated synergistic retrieval of water vapour profiles, to probe the lower parts of the atmosphere, which was previously largely unexplored, combining the near- and thermal infrared channels of SPICAM and PFS using nadir observations. Total abundances of water vapour can now be derived independently from different spectral ranges of PFS, NOMAD and ACS instruments and a combination of all absorption bands will be promising to improve the retrieval of its vertical profile from a nadir viewing geometry.

The synergy of different spectral absorption bands can be also applied to the retrievals of aerosols and many minor species. The new NOMAD and ACS solar occultation profiles can be combined with PFS nadir observations, which provide total dust, water ice contents and atmospheric temperatures up to 50 km. This combination can be applied as an input to the retrieval algorithms for many minor species. In addition, the comparison of PFS and TIRVIM atmospheric temperatures can help to verify the retrievals derived from the solar occultation channel of NOMAD and ACS in the low atmosphere. Another useful example is the collaboration for non-Local Thermodynamical Equilibrium (non-LTE) emission of $CO_2$ at 4.3 μm with the MEX/OMEGA infrared long channel, which can be compared with the same emission by TGO/ACS/TIRVIM, or TGO/NOMAD/LNO at 2.7 μm as described in [Lopez-Valverde, 2018]. Similar collaborations and cross-calibrations can also be done for ozone, dayglow, nightglows and many other interesting emissions in the infrared, visible and ultraviolet. More details are available in the references in Table 1.

| | **Mars Express** | **Trace Gas Orbiter** |
|---|---|---|

| | OMEGA<br><br>VNIR (0.38-1.05 µm)<br>IR-C* (0.93-2.73 µm)<br>IR-L (2.55-5.1 µm) | SPICAM<br><br>UV*(118-320 nm)<br>IR (1.1-1.7 µm) | PFS<br><br>SW (1.2-5.0µm)<br>LW (5-45 µm) | NOMAD<br><br>UVIS (200-650 nm)<br>LNO (2.3-3.8 µm)<br>SO (2.3-4.3 µm) | ACS<br><br>NIR (0.7-1.6 µm)<br>MIR (2.2-4.4 µm)<br>TIRVIM (1.7-17 µm) |
|---|---|---|---|---|---|
| **Temperature** | 5 µm<br><br>[Audouard, 2014] | 120-190 nm<br><br>[Forget, 2009] | 15 µm<br><br>[Grassi, 2005] | 2.7~3.7µm<br><br>[Vandaele, 2018, Mahieux 2012] | 15 µm<br>2.7~3.7µm<br><br>[Ignatiev, 2018; Guerlet, 2018; Korablev 2018] |
| **Dust** | 2.7 µm<br><br>[Vincendon, 2008] | 220-290 nm *<br>1–1.7 µm<br><br>[Mateshvili, 2007; Fedorova, 2009 & 2014] | 9.3 µm<br><br>[Wolkenberg, 2018] | 200-650 nm<br>2.3-4.0 µm<br><br>[Vandaele, 2019] | 9.3 µm<br>0.7~1.7 µm,<br>2.3–4.0 µm<br><br>[Ignatiev, 2019; Guerlet, 2019; Luginin, 2019] |
| **$CO_2$ ice clouds** | 2.7µm<br>4.26~4.3 µm<br><br>[Montmessin, 2007; Vincendon, 2011] | 100-200 nm *<br><br>[Montmessin 2006] | 2.7µm,<br>4.26~4.3 µm<br><br>[Aoki, 2018; Lopez-Valverde, 2018] | 2.7µm<br><br>[Vandaele, 2018; Lopez-Valverde, 2018] | 2.7µm,<br>4.26~4.3 µm, 10 µm<br><br>[Ignatiev 2018; Lopez-Valverde, 2018] |
| **Water ice clouds** | 3.1 µm<br><br>[Madeleine, 2012; Vincendon, 2011] | 300–320 nm *<br><br>[Mateshvili, 2009] | 12 µm<br><br>[Giuranna, 2019] | 300-320 nm<br>2.3–3.7 µm<br><br>[Vandaele, 2018] | 12 µm<br><br>[Ignatiev, 2019; Luginin, 2019] |
| **Water vapour** | 2.6 µm<br><br>[Melchiorri, 2006] | 1.38 µm<br><br>[Fedorova 2006; Montmessin 2018] | 32 – 20 µm<br><br>[Giuranna 2019; Montmessin 2018] | 2.55 µm, 3.3 µm, …<br><br>[Vandaele, 2018 & 2019] | 1.38 µm, 2.55 µm, 3.3 µm, …<br><br>[Korablev, 2018; Vandaele, 2019] |
| **CO** | 2.3 µm, 4.7 µm<br><br>[Encrenaz, 2006] | 190-270 nm *<br><br>[Cox, 2009] | 4.55 ~ 5.0 µm<br><br>[Billebaud, 2009] | 2.3 µm<br><br>[Smith, 2018; Vandaele, 2018] | 2.3 µm, 4.7 µm<br><br>[Smith, 2018; Korablev, 2018] |
| **Methane** | - | - | 3.3 µm<br><br>[Giuranna, 2019] | 3.27 ~ 3.3 µm<br><br>[Liuzzi. 2019; Korablev, 2019] | 3.27 ~ 3.3 µm<br><br>[Korablev, 2019] |

**Table 1. Summary list of the main atmospheric parameters that can be measured by the remote sensing spectrometers in MEX and TGO, including the most significant wavelengths and the corresponding references. *Note the OMEGA-IR-C channel and SPICAM-UV channel are no longer operational.**

Note that we have only discussed here the information for the spectrometers where the data is directly comparable. However, we must also mention the rest of the instruments that may also provide useful information for direct or indirect comparison of the atmospheric phenomena. First, radio science can provide very useful vertical profiles through the radio occultations, for which we have dedicated section 9.3. Then we have the imaging cameras (MEX/HRSC, MEX/VMC and TGO/CASSIS) that can provide very useful contextual information to the atmospheric measurements, in particular regarding dust and cloud coverage, and support the geological interpretation for any potential gas sources at the surface. Vice versa, the atmospheric instruments may provide important inputs for atmospheric correction of the images and improve the radiometric calibration of surface data. We shall also remark the potential synergies in the study of the upper atmosphere with MEX/MARSIS ionospheric sounding, which provides vertical electron density profiles and total electron content, and the MEX/ASPERA plasma measurements that provide an estimate of the atmospheric escape rate. Finally, we mention the subsurface information provided by the MEX/MARSIS and TGO/FREND, which may contribute to the understanding of potential underground volatiles.

## 5.2. Types of coordinated observation opportunities

In the following sections, we describe the science opportunity analysis for the two main types of combined observations, nadir and solar occultation measurements. In order to identify various possibilities, we first define here some basic terminology used for the various types of combined observations:

- **Simultaneous co-located observations:** These are observations by both spacecraft that occur exactly at the same time and cover the same location, therefore having the same illumination and local time conditions at the target, even if the areas covered by the field of views may not be exactly the same due to observing geometry. See example illustrations of a simultaneous solar occultation by both missions in Figure 3 and Figure 4, where we have simulated the orbits and positions of the two spacecraft using 3D visualization software Cosmographia, provided by NAIF-JPL and the ESA SPICE Service [Acton, 2018; Costa, 2018]. These simultaneous observations are the most scientifically interesting and are very useful for cross-comparison and cross-calibration between the different instruments. However, these "time-driven" opportunities can be extremely rare due to the limitations in the orbital evolution of both satellites and due to operational constraints. In the following sections we will describe and identify simultaneous observations using the same pointing mode (e.g. two solar occultations or two nadirs) although we will also mention other advanced future opportunities combining solar occultation and nadir.

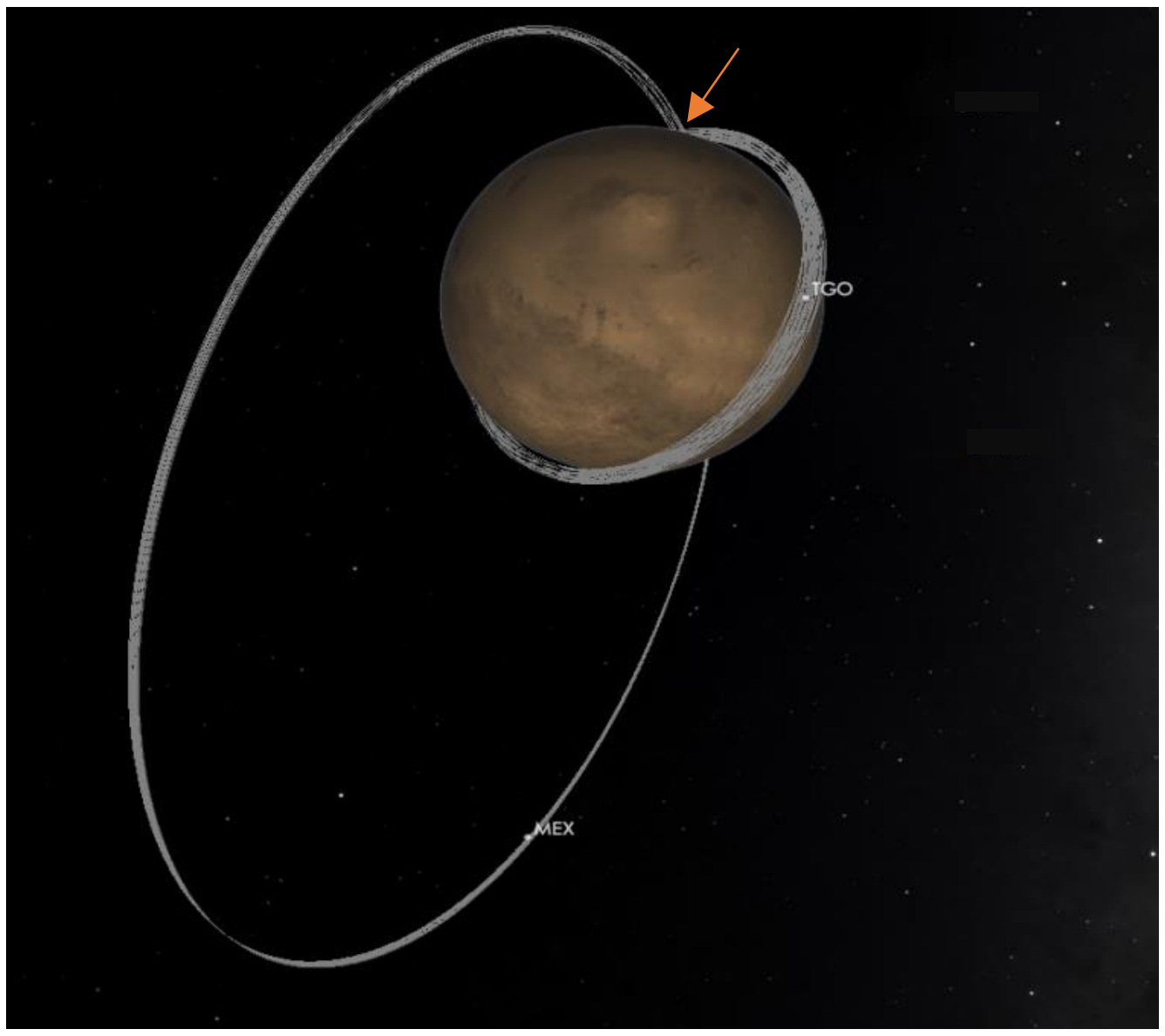

**Figure 3. 3D simulation of the MEX and TGO orbits as seen from the Sun, note the solar occultation point (arrow) occurring at similar latitude and longitude.**

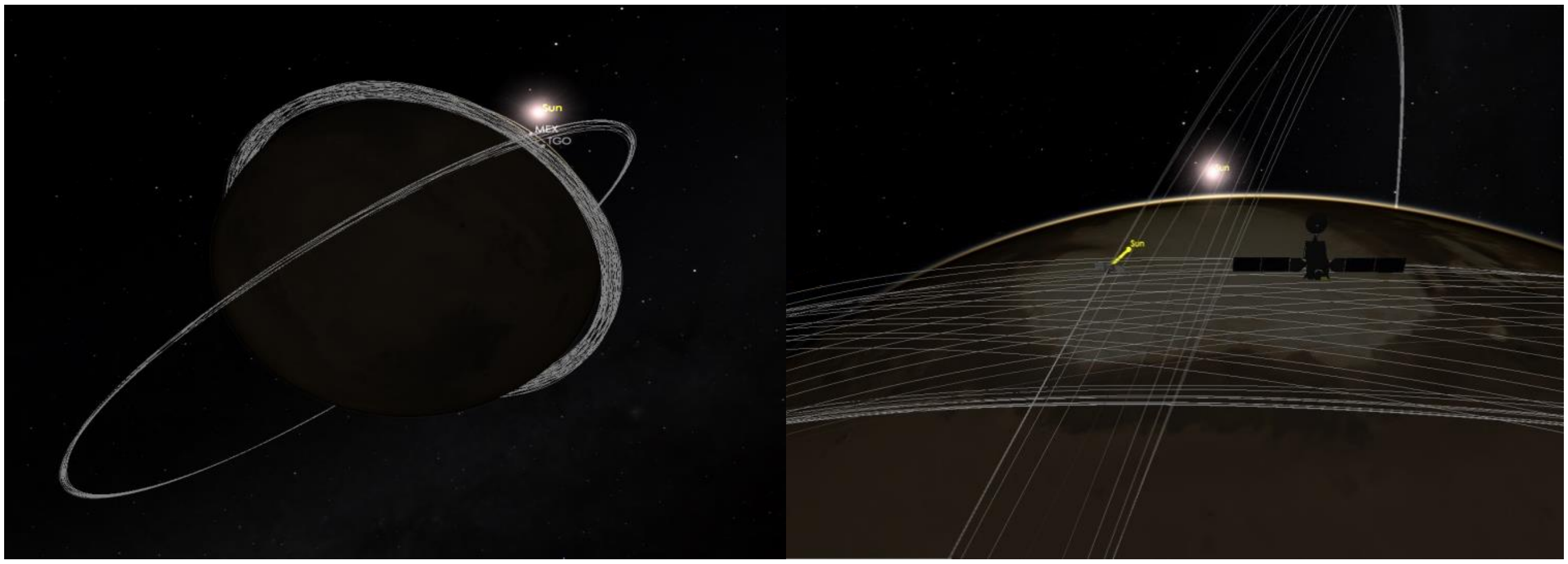

**Figure 4**. **3D simulation with SPICE Cosmographia of MEX and TGO simultaneous solar occultation observation at the same time, latitude and longitude.**

- **Quasi-simultaneous co-located observations:** These are observations that are not exactly simultaneous but chosen to observe over the same areas within certain margins in time, latitude and longitude coordinates. These observations are still “spatially driven” and the number of opportunities highly depends on the flexibility of the scientific requirements, which can be relaxed in terms of time and/or spatial differences between measurements depending on the science objective. This mode is extremely interesting to increase the statistics over target areas, as the data can provide very important information for broad scientific objectives linked to surface and/or atmospheric studies. We distinguish here two main types of quasi-simultaneous observations:

    - Surface-driven: These are observations of the same latitude and longitude coordinates, but performed at different times. They are useful for specific scientific objectives, in particular in relation with surface features and in general using the imaging cameras or spectrometers for which the comparison of the results can still be relevant even if observations are taken at very different times, as long as the illumination conditions at the surface are similar. For dynamic processes related to the atmosphere the time is more restrictive, but it can still be interesting to observe the same area after several minutes or even hours, considering that the illumination conditions vary typically ~15 deg/hour at the equator, and therefore depending on the scientific goal, combined observations may still be useful after a few hours.

    - Sun-illumination-driven: These are observations covering the same latitude region and with the same illumination conditions, but the longitude coordinates may be different. In other words, these are two observations of the same geometry on Mars with respect to the Sun, but for which the actual point on the surface is different. This is especially applicable to atmospheric dynamical features (like traveling waves and tides) strongly associated to the local time and high-atmosphere observations, which do not have a strong dependence on the surface features, but are much more dependent on the illumination conditions. The time differences in this case cause a longitudinal variation of ~15 deg/h (~900 km/h, ~15 km/min at equator).

- **Non-simultaneous seasonal observations:** These are observations of the same target area that do not occur close in time, but are performed within the same season with similar conditions and therefore can still be used to infer useful information. This is in particular applicable to surface geomorphology or mineralogy observations that are not expected to change within short timescales and therefore can be compared even if separated by several months or even years. Also regarding the atmosphere, these observations may serve many studies of climatological variability. In general the overall scientific requirement on both missions is to reach as much as possible a full coverage of the parameter space, not only in terms of latitude and longitude, but also in terms of season (solar longitude) and illumination conditions (solar elevation angle and local time) so that all the data can be ingested into the climate models for comparison. These non-simultaneous observations are of course much easier to obtain and are of frequent occurrence, limited only by the geometry of each orbit. In particular, it is limited by the pericentre evolution of the MEX orbit as not all latitudes may be seen within all possible conditions, or by the latitude limitation of TGO around the polar areas.

To perform the opportunity analysis we calculated possibilities of both nadir and occultation observations. All computations for this paper were performed using the MAPPS planning software and the SOLab tool based on SPICE [van der Plas, 2016; Costa, 2012] and based on the Long Term Planning (LTP) reference trajectories of Mars Express and the Trace Gas Orbiter, as generated by the Flight Dynamics teams and processed for both missions by the Science Operations Centres and the ESA SPICE Service [Costa, 2018]. Note the spacecraft trajectories used here are the predicted ephemerides used for long-term planning and therefore some small differences may appear due to the spacecraft phasing error along the orbit with respect to the final measured or reconstructed trajectories. In any case this time phasing is controlled daily by Flight Dynamics to stay within a small threshold (+/-10s, applicable for both missions) and it does not imply a significant impact to the calculations done for this study.

The full list of reference trajectories and ephemeris used in this paper are given in the Data Availability section at the end of this paper.

## 6. Analysis of coordinated solar occultations

As previously mentioned, the main scientific goal of the ExoMars TGO mission is the analysis of trace gas species. Solar Occultation, illustrated in Figure 5, is the most important pointing mode to cover the main requirements of the NOMAD and ACS spectrometers that will provide the high-resolution vertical profiles of various gasses and their densities [Vandaele, 2019 and Korablev, 2019]. Mars Express carries the SPICAM instrument that is also able to perform solar occultation measurements in the infrared range. Both spacecraft also make measurements to allow vertical thermal profiles to be determined. Comparison of the results from both missions is of crucial importance for verification of vertical profiles and the retrieval methods. [Montmessin, 2017]

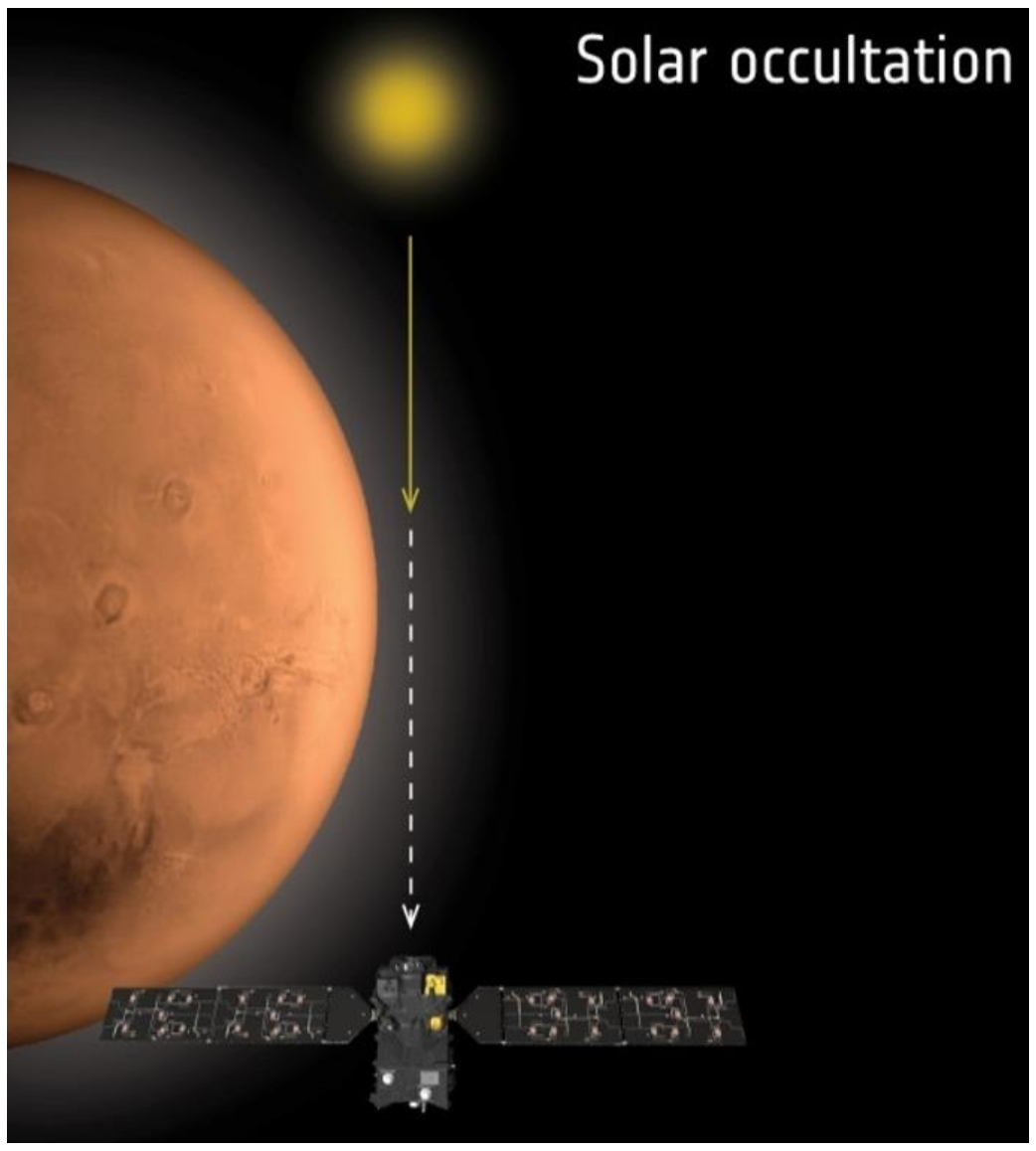

**Figure 5. Illustration of the solar occultation method.**
**Solar radiation is absorbed by the planet's atmosphere.**

In this section we have performed an analysis of all occultation points MEX-Mars-Sun and TGO-Mars-Sun for the two observing directions: ingress (sunset, observation starting outside the atmosphere and moving down towards the surface) and egress (sunrise, observation starting at the surface level and moving up outside the atmosphere). The occultation point is defined as the intercept between the spacecraft and Sun centre vector with the Mars reference ellipsoid (equatorial and polar radius 3389.50 and 3396.19 as per [Archinal, 2010]), at a limb height of 0 km with respect to this ellipsoid.

We note that the solar occultation measurements cover a wide range of spatial coordinates depending on the altitude profile, starting outside the atmosphere (around 200~250 km) down to the surface (at 0 km) for the ingress and vice versa for the egress. The actual occultation profile is highly dependent on the spacecraft orbit geometry, in particular the beta angle, and the detailed start/end times can only be confirmed during the final scheduling of the observations, since they are driven by the specific scientific objectives and the seasonal atmospheric conditions. After discussion with the instrument teams, it was agreed to use the 0 km reference altitude for the scope of this long term opportunity analysis, since it simplifies the computations and provides sufficient accuracy to identify the best opportunities within the wide filter ranges that we will discuss further below.

The plots in Figure 6 and Figure 7 show the spatial coverage (latitude, longitude) of the Mars Express and the ExoMars Trace Gas Orbiter solar occultations as computed for the TGO-Sun vector crossing at a limb height of 0 km, using the Mars reference ellipsoid. The plots correspond to about 1 year.

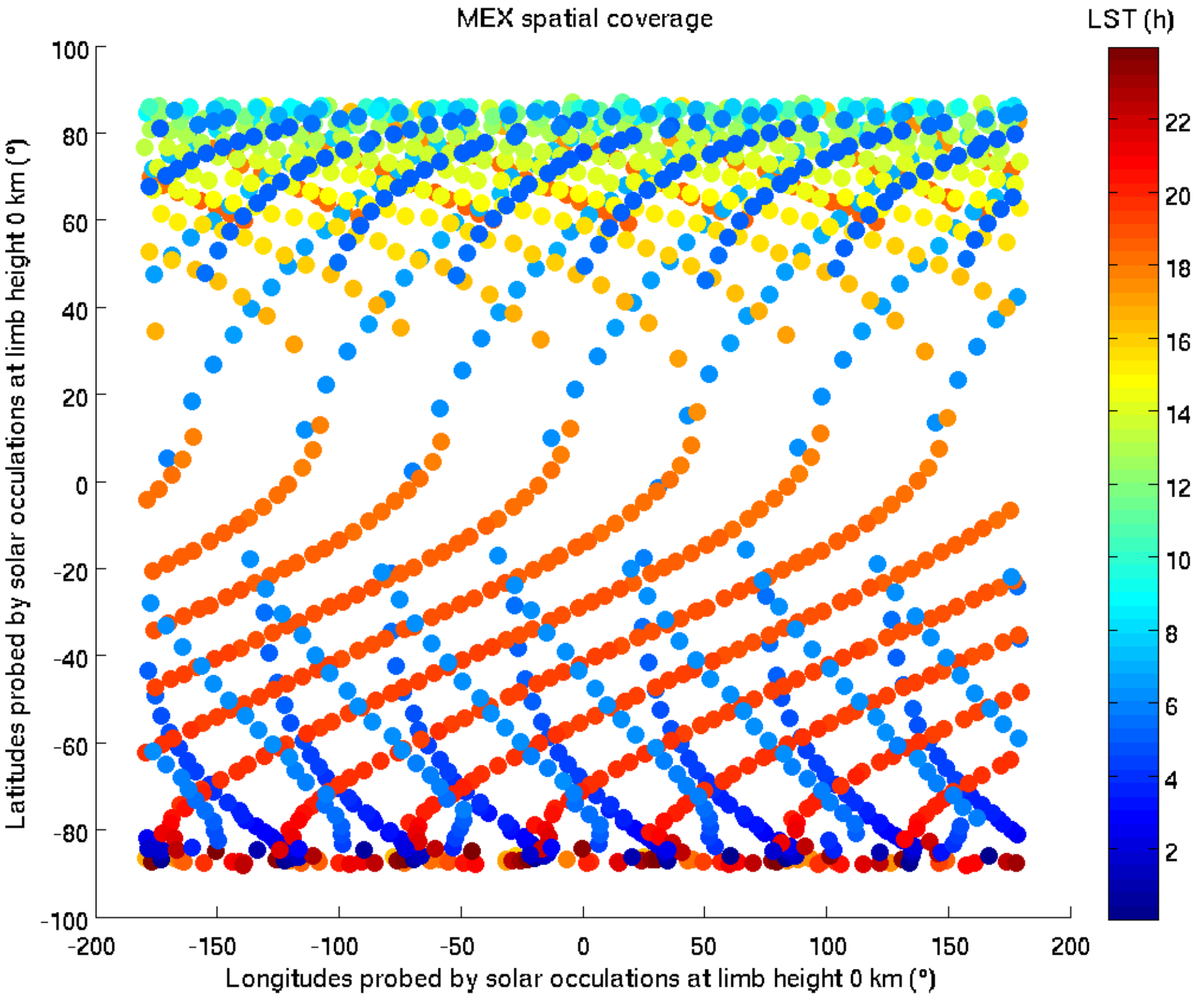

**Figure 6**. **Mars Express solar occultations from 21 April 2018 until 3 April 2019, coloured by local solar time for latitude and longitude spatial coverage.**

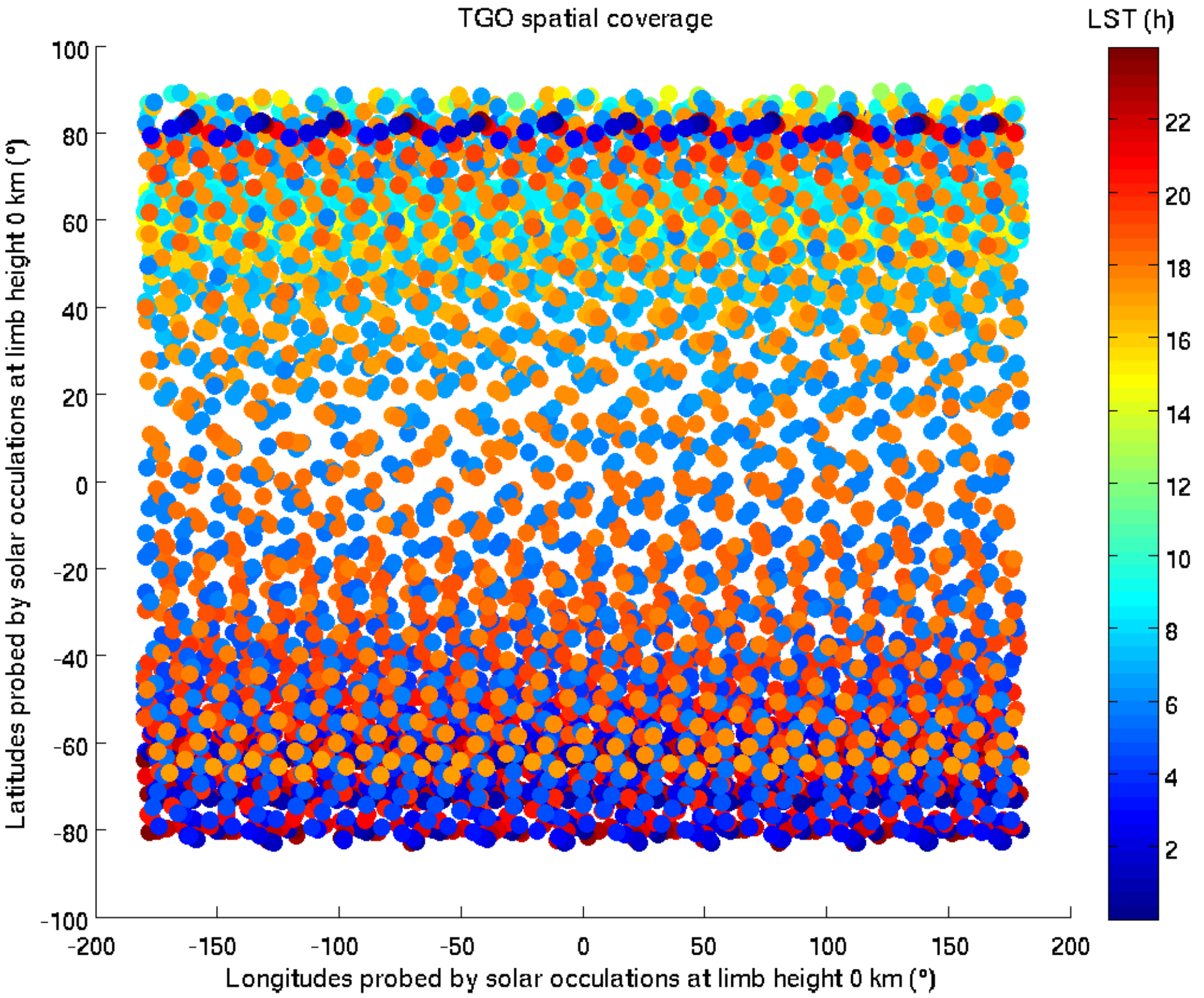

**Figure 7**. **Trace Gas Orbiter solar occultations from start of the nominal TGO science phase on 21 April 2018 until 3 April 2019, coloured by local solar time for latitude and longitude spatial coverage.**

The solar occultation measurements for both missions are combined in Figure **8** with respect to the Martian year. These latitudinal profiles allow us to identify the seasons where joint solar occultation measurements are possible. As previously mentioned, the ideal case for a simultaneous co-located observation would be to observe the same latitude and longitude at the same time. We can see from these profiles that joint observations can only occur in very specific occasions when the latitudinal trends from both missions do actually cross. Although we see that the latitude profiles often overlap, in some cases we can observe that the local times are not the same, meaning that the two spacecraft are actually viewing the same latitude but one satellite is observing the dusk terminator while the other is probing the dawn terminator.

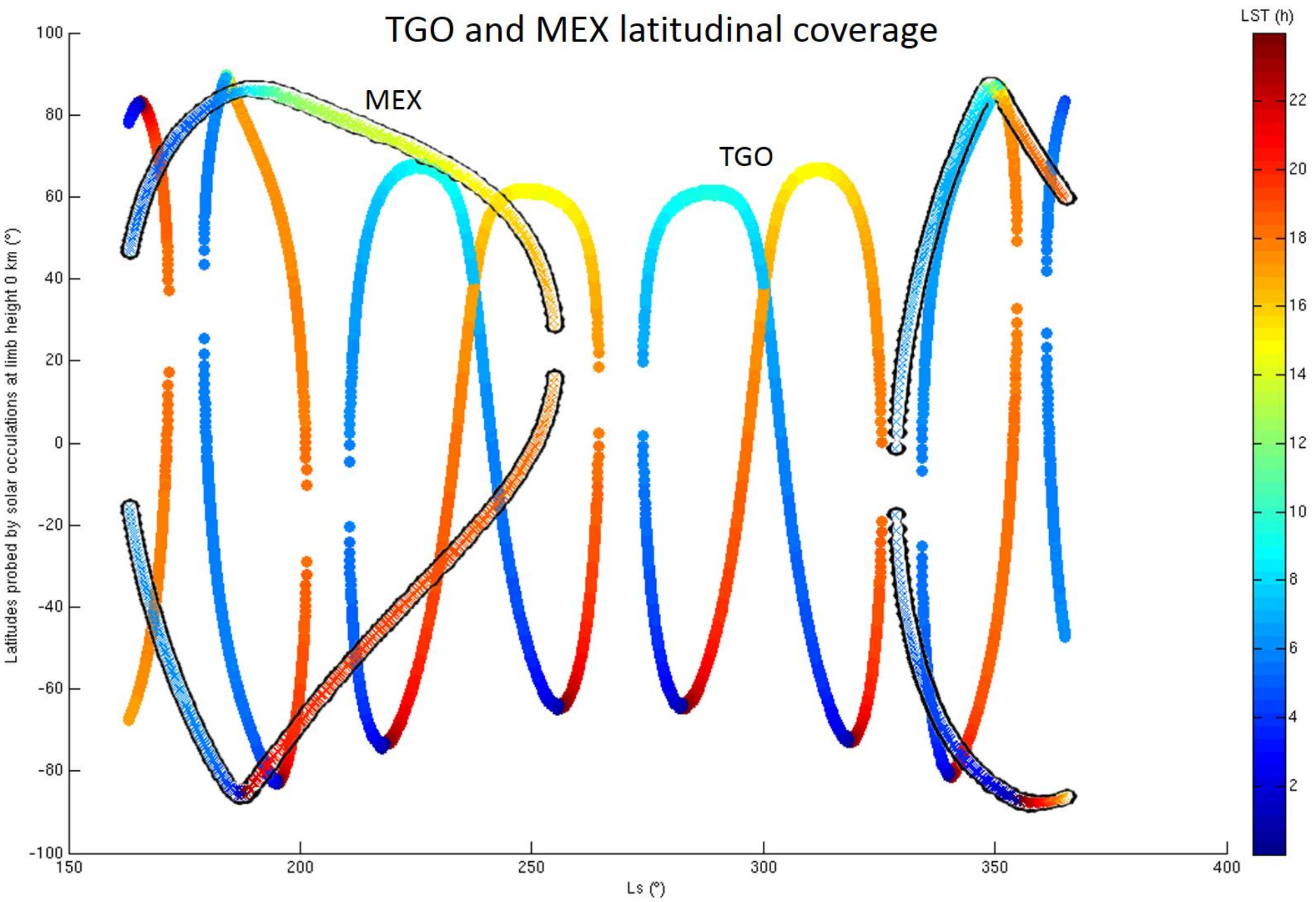

**Figure 8**. **Latitude and local time of both TGO (circles) and MEX (crosses with contour lines) solar occultations from 21 April 2018 until 3 April 2019, coloured by local solar time and distributed over latitude and seasonal evolution over the Martian year (Ls= Solar Longitude).**

We see that the occurrence of simultaneous observations is very limited and we only find a couple of joint opportunities within 1 minute time difference but not exactly at the same location, as shown in Table 2. Therefore we can now consider the quasi-simultaneous case and extend the time, latitude and longitude ranges, that is, to consider observations that are not exactly co-located but are still in the same region within a certain time margin. In this case we can increase the number of joint occultations that can still cover most of the scientific requirements for coordinated atmospheric studies. A few minutes and a few degrees difference in latitude or longitude can still be of high interest for comparison of the vertical profiles, especially when looking at the upper atmospheric layers.

In order to identify the quasi-simultaneous opportunities, we first computed all solar occultation events for both missions and then compared all times and geometrical conditions using an extended range of filters. Various filtering parameters have been discussed and used during the long term planning process. The following set of filtering parameters have been agreed between the relevant instrument teams (MEX/SPICAM, TGO/NOMAD and TGO/ACS) for the differences between solar occultations measured by both missions:

- Time difference between 2~60 minutes
- Latitude and longitude difference between the occultation points 5 ~ 30 degrees
- Distance between the occultation points 200 ~ 2000 km

Note that the narrow filters have been used to identify the top priority joint observations, but many other filtering parameters have been used during the planning process, with larger margins to maximize the number of joint opportunities and also to prevent any possible uncertainty in the geometry. Note in particular that for the polar regions, we typically filter based on the distance between the occultation points instead of the longitude to avoid problems with the singularity at the poles. Also, as previously mentioned, the importance of temporal or spatial differences is highly dependent on the scientific objective, and therefore some observations may still be scientifically relevant even if taken with longer time or larger spatial separation (e.g. for upper atmosphere measurements).

The current long-term trajectories of both missions are confirmed as a valid reference until January 2021. We therefore obtained long lists of joint quasi-simultaneous opportunities for the whole mission up to December 2020 and these were made available to the planning teams of both missions. The complete lists of opportunities, partially shown hereafter, is also available in the supplementary information, also including other various combined filtering parameters for past and future periods.

Our analysis confirmed that MEX-TGO combined occultations are geometrically possible and they can be planned regularly encompassing several days or weeks. We identified many quasi-simultaneous occultations that can be observed within less than 30 degrees latitude and longitude and few minutes from each other. In particular, we observed that the periods May-June and August-September 2018, and May-June 2019 exhibit many combined occultation events.

The joint opportunity events for 2018 and 2019 were provided as a high priority input during the planning process and finally a down-selection was performed by the teams based on the scientific requirements and the operational constraints. Table 2 shows all the joint solar occultation observations that were finally selected, planned and executed between May 2018 and June 2019. We can see in particular a couple of observations, 29 August 2018 and 15 June 2019, where the time difference is less than one minute and simultaneous observations of the two missions have been planned. The type of occultation event (ingress or egress, sunset or sunrise) is also provided to facilitate the interpretation of the data since it indicates the direction of the vertical profile obtained (downwards or upwards). Note that the list of other joint opportunities that were identified but not planned for operational reasons is also available in the supplementary information.

| Reference Occultation Points | | | Differences between Occultations | | | | Observing Instruments | | | |
|---|---|---|---|---|---|---|---|---|---|---|
| Date Time (UTC) | Lat. (deg) | Lon. (deg) | Time (mm:ss) | Lat. (deg) | Lon. (deg) | Distance (km) | TGO | Event | MEX | Event |
| 21-05-2018T12:03:43 | 80 | 67 | 07:43 | 25 | -26 | 1692 | ACS | egress | SPICAM | egress |
| 24-05-2018T02:55:22 | 81 | 225 | 12:52 | 7 | -25 | 532 | NOMAD | ingress | SPICAM | egress |
| 25-05-2018T14:02:40 | 80 | 104 | 11:55 | **-2** | 26 | 209 | ACS | egress | SPICAM | egress |
| 26-05-2018T03:48:53 | 82 | 233 | **01:33** | **0** | -20 | **135** | NOMAD | ingress | SPICAM | egress |
| 28-05-2018T04:52:50 | 87 | 268 | 10:26 | **4** | 20 | **143** | NOMAD | ingress | SPICAM | egress |
| 21-06-2018T14:53:40 | -77 | 219 | 05:32 | -6 | 10 | 366 | NOMAD | egress | SPICAM | ingress |
| 24-06-2018T05:43:20 | -70 | 1 | 03:06 | **-2** | -10 | **149** | NOMAD | egress | SPICAM | ingress |
| 24-06-2018T19:38:30 | -66 | 171 | 10:50 | **0** | 9 | **189** | ACS | egress | SPICAM | ingress |
| 26-06-2018T20:32:14 | -65 | 172 | **01:24** | -12 | 16 | 914 | ACS | egress | SPICAM | ingress |
| 03-08-2018T02:40:30 | -60 | 94 | 02:37 | -24 | 25 | 1699 | NOMAD | ingress | SPICAM | ingress |
| 08-08-2018T08:25:30 | -50 | 38 | 04:16 | -18 | **5** | 1020 | ACS | egress | SPICAM | ingress |
| 10-08-2018T23:18:50 | -41 | 197 | 06:08 | -10 | **3** | 586 | NOMAD | ingress | SPICAM | ingress |
| 14-08-2018T04:02:37 | -29 | 150 | 04:17 | **0** | **1** | **48** | ACS | egress | SPICAM | ingress |
| 16-08-2018T05:07:20 | -20 | 147 | 11:14 | 8 | -10 | 665 | ACS | egress | SPICAM | ingress |
| 19-08-2018T09:49:00 | -1 | 103 | 02:57 | 23 | -13 | 1596 | NOMAD | ingress | SPICAM | ingress |
| 26-08-2018T16:51:40 | 41 | 45 | 08:11 | -25 | 17 | 1436 | ACS | egress | SPICAM | egress |
| 29-08-2018T21:32:19 | 64 | 349 | **00:19** | 11 | -9 | 610 | ACS | egress | SPICAM | egress |
| 04-09-2018T03:18:20 | 60 | 311 | 04:18 | **0** | -9 | 234 | ACS | egress | SPICAM | egress |
| 04-09-2018T17:12:01 | 60 | 116 | 08:21 | **-2** | **4** | 231 | NOMAD | ingress | SPICAM | egress |
| 07-09-2018T08:02:44 | 58 | 282 | 06:44 | -6 | 11 | 506 | NOMAD | ingress | SPICAM | egress |
| 09-09-2018T22:53:45 | 54 | 84 | 05:45 | -9 | 14 | 741 | ACS | egress | SPICAM | egress |
| 12-09-2018T13:44:25 | 52 | 251 | 04:35 | -9 | 24 | 971 | NOMAD | ingress | SPICAM | merged |
| 15-09-2018T04:34:51 | 48 | 55 | 03:01 | -15 | 23 | 1217 | NOMAD | ingress | SPICAM | merged |
| 21-04-2019T09:04:50 | 59 | -21 | 03:27 | 13 | **3** | 787 | ACS | ingress | SPICAM | ingress |
| 26-04-2019T14:51:26 | 43 | -63 | **01:09** | **0** | **2** | **95** | BOTH | ingress | SPICAM | ingress |
| 23-05-2019T09:48:52 | -77 | -147 | 05:11 | **2** | 8 | 150 | NOMAD | ingress | SPICAM | egress |
| 25-05-2019T10:47:13 | -75 | -132 | 08:12 | **1** | 10 | 168 | NOMAD | ingress | SPICAM | egress |
| 28-05-2019T15:32:04 | -75 | -178 | 04:19 | **1** | 5 | **93** | ACS | ingress | SPICAM | egress |
| 31-05-2019T06:23:41 | -73 | -13 | 03:53 | **0** | **2** | **52** | ACS | ingress | SPICAM | egress |
| 02-06-2019T21:15:20 | -72 | 150 | 03:27 | **0** | **2** | **44** | NOMAD | ingress | SPICAM | egress |
| 05-06-2019T12:07:00 | -71 | -45 | 03:10 | **1** | **0** | **34** | NOMAD | ingress | SPICAM | egress |
| 15-06-2019T23:35:11 | -61 | -108 | **00:17** | 5 | 6 | 355 | ACS | ingress | SPICAM | egress |
| 21-06-2019T05:17:35 | -64 | -147 | 07:03 | 13 | 9 | 813 | NOMAD | ingress | SPICAM | egress |
| 26-06-2019T03:41:31 | 20 | -38 | 02:10 | 8 | **3** | 813 | NOMAD | egress | SPICAM | ingress |

**Table 2. List of planned observations for TGO and MEX quasi-simultaneous solar occultation from May 2018 to June 2019. The reference times (in universal time, UTC) and geometric parameters correspond to the last occultation of the pair. The differences correspond to the separation between the two occultation points in**

**time and spatial coordinates (latitude, longitude and distance). The last columns reflect the instruments that were observing in each of the solar occultations, either ingress (start of the occultation, or sunset from the point of view of the spacecraft), egress (end of the occultation or sunrise as seen from the spacecraft) or merged (both).**

The next Mars Express eclipse season starts at the end of 2019 and as a complement for the preparation of future science observation, we now provide the full list of joint solar occultation opportunities from December 2019 until December 2020 using the same filtering parameters above. These opportunities will again be used by the planning teams to accommodate the observations following the scientific priorities and operational constraints in the coming year.

| Reference Occultation Points | | | Differences | | | |
|---|---|---|---|---|---|---|
| Date Time (UTC) | Lat. (deg) | Lon. (deg) | Time (mm:ss) | Lat. (deg) | Lon. (deg) | Dist. (km) |
| 22-12-2019T01:11:23 | -34 | 95 | 04:11 | 23 | 15 | 1894 |
| 22-12-2019T14:57:40 | -32 | -108 | 14:15 | 17 | 15 | 1477 |
| 24-12-2019T16:01:37 | -24 | -112 | 07:38 | **4** | **1** | 1458 |
| 25-12-2019T05:55:39 | -16 | 46 | 03:19 | 10 | **1** | 1445 |
| 25-12-2019T19:42:05 | -13 | -156 | 14:24 | 15 | **2** | 1064 |
| 31-12-2019T09:17:02 | 47 | 21 | 12:06 | 29 | 18 | 1068 |
| 31-12-2019T23:15:09 | 20 | -163 | **00:26** | 28 | 21 | 729 |
| 01-01-2020T13:01:19 | 23 | -5 | 12:58 | 26 | 24 | 738 |
| 03-01-2020T00:10:31 | 51 | 179 | 14:50 | 22 | 16 | 431 |
| 03-01-2020T14:09:02 | 52 | -25 | 02:18 | 21 | 19 | 448 |
| 04-01-2020T03:57:22 | 33 | 151 | 10:13 | 19 | 22 | 165 |
| 06-01-2020T05:02:03 | 55 | 133 | 04:45 | 15 | 16 | 188 |
| 06-01-2020T18:52:44 | 42 | -53 | 07:47 | 13 | 19 | 212 |
| 08-01-2020T19:54:47 | 57 | -69 | 07:12 | 9 | 13 | **91** |
| 09-01-2020T09:47:41 | 50 | 103 | 05:28 | 8 | 16 | **94** |
| 11-01-2020T10:47:14 | 60 | 90 | 09:13 | **5** | 9 | **116** |
| 12-01-2020T00:42:09 | 56 | -103 | 03:19 | **4** | 12 | 339 |
| 14-01-2020T01:39:32 | 62 | -111 | 11:14 | **2** | **4** | 332 |
| 14-01-2020T15:36:22 | 61 | 50 | **01:26** | **1** | 6 | 323 |
| 15-01-2020T05:21:56 | 62 | -153 | 14:07 | **1** | 8 | 589 |
| 16-01-2020T16:31:45 | 64 | 47 | 13:06 | **1** | **2** | 588 |
| 17-01-2020T06:30:00 | 64 | -158 | **00:26** | **1** | **1** | 377 |
| 17-01-2020T20:15:53 | 66 | -3 | 12:14 | **2** | **0** | **132** |
| 19-01-2020T07:23:48 | 65 | -154 | 14:41 | **3** | 13 | **102** |
| 19-01-2020T21:22:01 | 66 | 1 | **02:01** | **3** | 12 | **140** |
| 20-01-2020T11:09:28 | 69 | 144 | 10:48 | **3** | 12 | **162** |
| 22-01-2020T12:14:00 | 67 | 160 | 03:27 | **3** | 26 | 399 |
| 23-01-2020T02:02:45 | 70 | -72 | 09:30 | **3** | 26 | 427 |
| 07-03-2020T07:36:58 | 79 | -98 | 11:31 | **2** | 29 | 670 |
| 09-03-2020T08:41:40 | 77 | -124 | 02:53 | **2** | 7 | 702 |
| 09-03-2020T22:30:00 | 78 | 38 | 09:39 | **1** | 7 | 962 |
| 11-03-2020T23:32:51 | 77 | 34 | 05:02 | **1** | 9 | 1004 |
| 12-03-2020T13:23:15 | 75 | -179 | 07:38 | **2** | 8 | 1048 |
| 14-03-2020T14:23:59 | 77 | -168 | 07:12 | **5** | 18 | 1350 |
| 15-03-2020T04:16:43 | 71 | -28 | 05:20 | 6 | 16 | 1413 |
| 17-03-2020T05:15:21 | 77 | -9 | 09:39 | 9 | 22 | 1775 |
| 17-03-2020T19:10:34 | 66 | 126 | 02:53 | 10 | 19 | 1682 |
| 19-03-2020T20:06:37 | 76 | 148 | 12:32 | 14 | 23 | 1894 |
| 20-03-2020T10:04:37 | 76 | -57 | **00:05** | 16 | 20 | 1477 |
| 20-03-2020T23:50:19 | 60 | 81 | 12:23 | 16 | 18 | 1458 |
| 23-03-2020T00:55:55 | 76 | 103 | 03:19 | 21 | 22 | 1445 |
| 23-03-2020T14:44:51 | 53 | -121 | 09:04 | 23 | 17 | 1064 |
| 25-03-2020T15:47:15 | 75 | -100 | 06:55 | 29 | 21 | 1068 |
| 31-03-2020T17:54:14 | -77 | -94 | 05:11 | 28 | **3** | 729 |

| Reference Occultation Points | | | Differences | | | |
|---|---|---|---|---|---|---|
| Date Time (UTC) | Lat. (deg) | Lon. (deg) | Time (mm:ss) | Lat. (deg) | Lon. (deg) | Dist. (km) |
| 02-04-2020T18:52:58 | -48 | -90 | 06:37 | 19 | **2** | 1151 |
| 03-04-2020T08:43:49 | -62 | 67 | 07:21 | 15 | **1** | 905 |
| 19-05-2020T15:25:33 | -16 | 54 | 04:11 | 30 | 8 | 1803 |
| 20-05-2020T05:15:33 | -43 | -140 | 08:12 | 28 | 10 | 1751 |
| 22-05-2020T06:17:24 | -13 | -144 | 08:04 | 22 | **4** | 1333 |
| 22-05-2020T20:11:34 | -33 | 18 | 04:02 | 20 | 6 | 1251 |
| 25-05-2020T11:07:51 | -9 | 174 | **00:52** | 10 | **2** | 579 |
| 26-05-2020T00:55:11 | -15 | -26 | 10:57 | 7 | **4** | 477 |
| 28-05-2020T02:00:12 | -5 | -24 | 07:03 | **5** | **3** | 313 |
| 28-05-2020T15:54:07 | 4 | 131 | 04:28 | 8 | **0** | 495 |
| 30-05-2020T16:59:00 | 26 | 130 | 08:47 | **3** | **2** | **199** |
| 31-05-2020T06:51:10 | 2 | -68 | 03:53 | 27 | 7 | 1664 |
| 31-05-2020T06:55:00 | 29 | -75 | 01:35 | **5** | **1** | 293 |
| 05-10-2020T23:26:08 | -23 | 9 | 04:02 | 20 | 11 | 1306 |
| 06-10-2020T13:16:01 | -45 | 155 | 10:22 | 19 | 8 | 1170 |
| 08-10-2020T00:25:51 | -32 | 8 | 13:24 | 17 | 14 | 1184 |
| 08-10-2020T14:25:00 | -50 | 151 | **00:26** | 16 | 12 | 1086 |
| 09-10-2020T04:10:39 | -51 | -52 | 14:15 | 16 | 9 | 1001 |
| 10-10-2020T15:23:08 | -40 | 164 | 10:31 | 15 | 17 | 1110 |
| 11-10-2020T05:19:19 | -56 | -56 | 03:01 | 15 | 15 | 1030 |
| 13-10-2020T06:19:02 | -45 | -40 | 08:47 | 14 | 22 | 1121 |
| 13-10-2020T20:13:15 | -60 | 95 | 04:45 | 14 | 20 | 1059 |
| 15-10-2020T21:14:02 | -50 | 116 | 07:29 | 13 | 29 | 1188 |
| 16-10-2020T11:06:54 | -63 | -116 | 05:54 | 12 | 27 | 1127 |
| 03-12-2020T04:59:59 | -69 | 26 | 14:33 | **4** | 28 | 591 |
| 03-12-2020T18:58:11 | -69 | -178 | 01:35 | **4** | 27 | 577 |
| 04-12-2020T08:45:07 | -73 | 3 | 11:14 | **4** | 26 | 559 |
| 06-12-2020T09:50:12 | -69 | -19 | 02:36 | **3** | 10 | 261 |
| 06-12-2020T23:38:16 | -71 | 146 | 10:13 | **2** | 10 | 244 |
| 09-12-2020T00:42:06 | -69 | 141 | 04:11 | **0** | **4** | **85** |
| 09-12-2020T14:31:47 | -68 | -66 | 08:30 | **1** | **2** | **63** |
| 11-12-2020T15:33:54 | -69 | -60 | 06:12 | **4** | 12 | 363 |
| 12-12-2020T05:25:31 | -64 | 85 | 06:29 | **5** | 11 | 379 |
| 14-12-2020T06:25:46 | -69 | 100 | 08:04 | 8 | 19 | 694 |
| 14-12-2020T20:19:27 | -59 | -122 | 04:36 | 9 | 17 | 717 |
| 16-12-2020T21:17:40 | -69 | -100 | 10:22 | 14 | 24 | 1063 |
| 17-12-2020T11:13:49 | -53 | 34 | 02:01 | 16 | 21 | 1099 |
| 18-12-2020T00:59:29 | -51 | -168 | 14:33 | 17 | 19 | 1149 |
| 19-12-2020T12:09:32 | -69 | 60 | 13:24 | 21 | 27 | 1492 |
| 20-12-2020T02:07:43 | -68 | -144 | **00:52** | 23 | 25 | 1551 |
| 20-12-2020T15:54:26 | -44 | -11 | 11:31 | 25 | 22 | 1624 |
| 29-12-2020T16:09:44 | 34 | 57 | 09:48 | 25 | **1** | 1476 |
| 30-12-2020T06:06:55 | 16 | -146 | 01:18 | 18 | **1** | 1042 |
| 30-12-2020T19:54:35 | 24 | 12 | 11:48 | 10 | **2** | 579 |

**Table 3. List of future opportunities for TGO and MEX quasi-simultaneous solar occultation from December 2019 to December 2020. The reference times and geometric parameters correspond to the last occultation of the pair. The differences correspond to the separation between the two occultation points in time and spatial coordinates (latitude/longitude).**

## 7. Analysis of coordinated Nadir Observations

Both the Mars Express and Trace Gas Orbiter satellites are also observing in nadir attitude performing high-resolution imaging of the surface with HRSC and CASSIS, respectively. Mars Express is also characterizing the surface composition with the OMEGA imaging spectrometer in nadir mode. Nadir geometry is also used by Mars Express for atmospheric studies, retrieving atmospheric temperature and various column abundances with the PFS and SPICAM spectrometers, performing cloud monitoring and acquiring overall contextual information with HRSC, OMEGA and the VMC camera. These same nadir observations are also used by other instruments such as FREND on-board TGO and the MARSIS radar on-board Mars Express to provide mapping information of the subsurface that is of great importance for the analysis of the ice content below the surface and other contextual interpretation of the data.

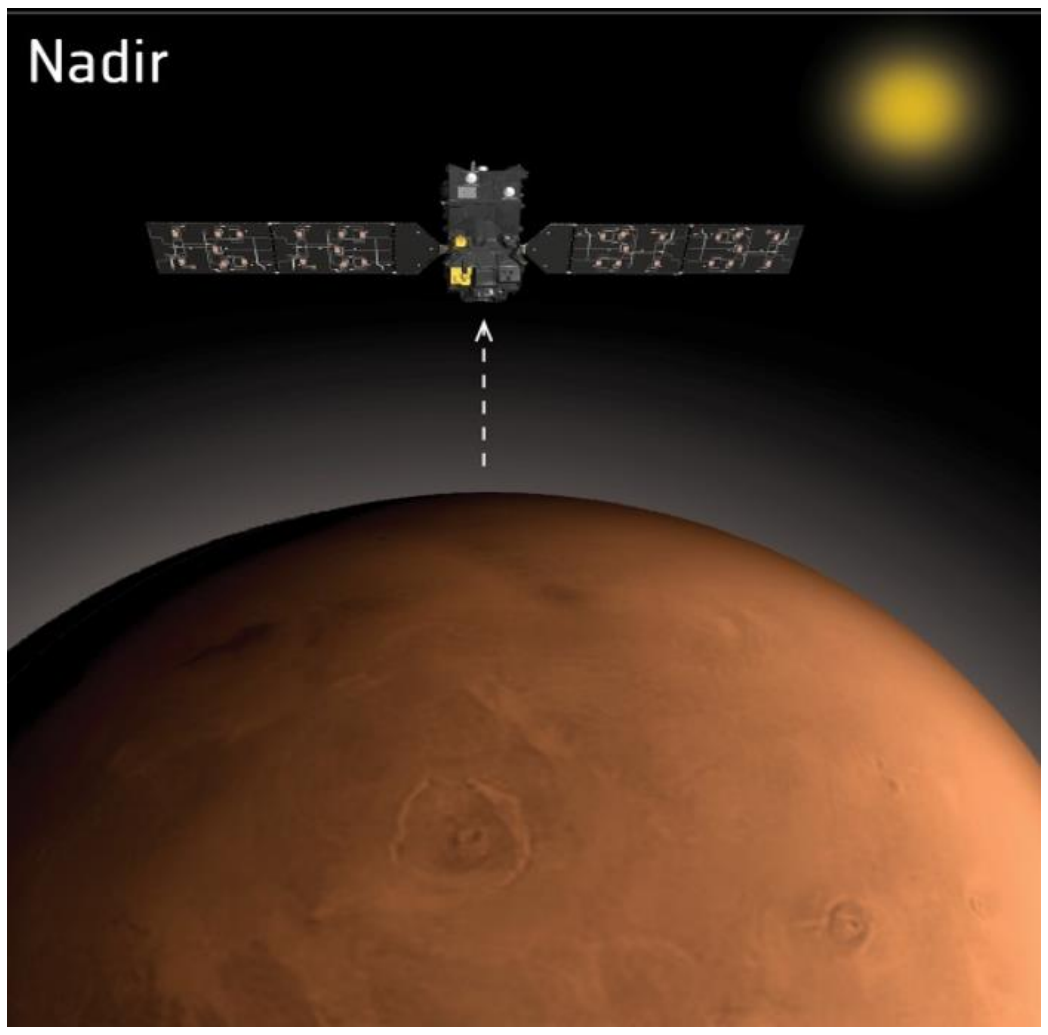

**Figure 9. Illustration of nadir observation pointing. The spacecraft observes signal emitted or reflected from the surface and the atmosphere of Mars.**

Among the wide range of scientific objectives that can be covered with nadir observations, some scientific targets are especially suited for combining data from the two missions. In particular, regarding atmospheric studies, we can remark the integrated columns of water vapour and aerosols (water ice and/or dust), and the thermal structure in the troposphere (see section 5.1 and Table 1). This makes a special focus on PFS and SPICAM on the MEX side, and on ACS and NOMAD on the TGO side. Just to give two specific examples: the ACS Thermal InfraRed spectrometer (TIRVIM) and PFS nadir observations can be combined to retrieve atmospheric temperatures, dust and water ice contents and surface temperatures, and the ACS Near InfraRed (NIR) channel, NOMAD UltraViolet and Visible channel (UVIS) and the SPICAM near-IR channel should be combined and compared to study the ozone profiles.

As discussed in section 2, MEX can cover the whole planet up to high latitudes (see Figure 10) and the science campaigns have a very flexible observing pattern, highly dependent on the orbital and seasonal variations both of the mission and the planet. However, the science observations are restricted by the communication passes, so the number of nadir observations is very limited compared to TGO, which can observe almost continuously in nadir mode except for the solar occultations and the operational limitations explained in Section 3. Fortunately, MEX has a lot of flexibility regarding the pointing modes and therefore it can accommodate science requests to observe together with TGO when needed, provided that the opportunities are identified in advance, which is the goal of this work. The HRSC, OMEGA and VMC imaging instruments typically observe the dayside of the planet, either at low altitude for high-resolution imaging, or at high altitude for global monitoring of the atmosphere. The MARSIS subsurface radar observes only at low altitudes during the nightside, and also in the dayside for ionosphere sounding. The rest of the instruments perform very variable observations at all altitudes and illumination conditions to maximize the overall atmospheric coverage, or the magnetosphere in the case of ASPERA.

The TGO instruments point towards nadir by default, only changed for solar occultation pointings and some special calibrations or operational manoeuvres. The observation pattern is therefore more regular compared to MEX, although it cannot reach high latitudes near the poles due to the 74º orbit inclination (see Figure 10). In particular the NOMAD instrument observes most dayside passes in nadir with two channels (LNO and UVIS), while the nightside observations are less frequent. In the case of the ACS instrument, the NIR channel acquires 20-min observations per orbit centred on the middle of the dayside pass, whereas the TIRVIM channel observes both day- and nightside, but with very variable duty cycle depending on operational restrictions. The FREND instrument observes continuously and CASSIS is target-oriented, observing only for special pre-selected landmarks in each orbit with good illumination conditions.

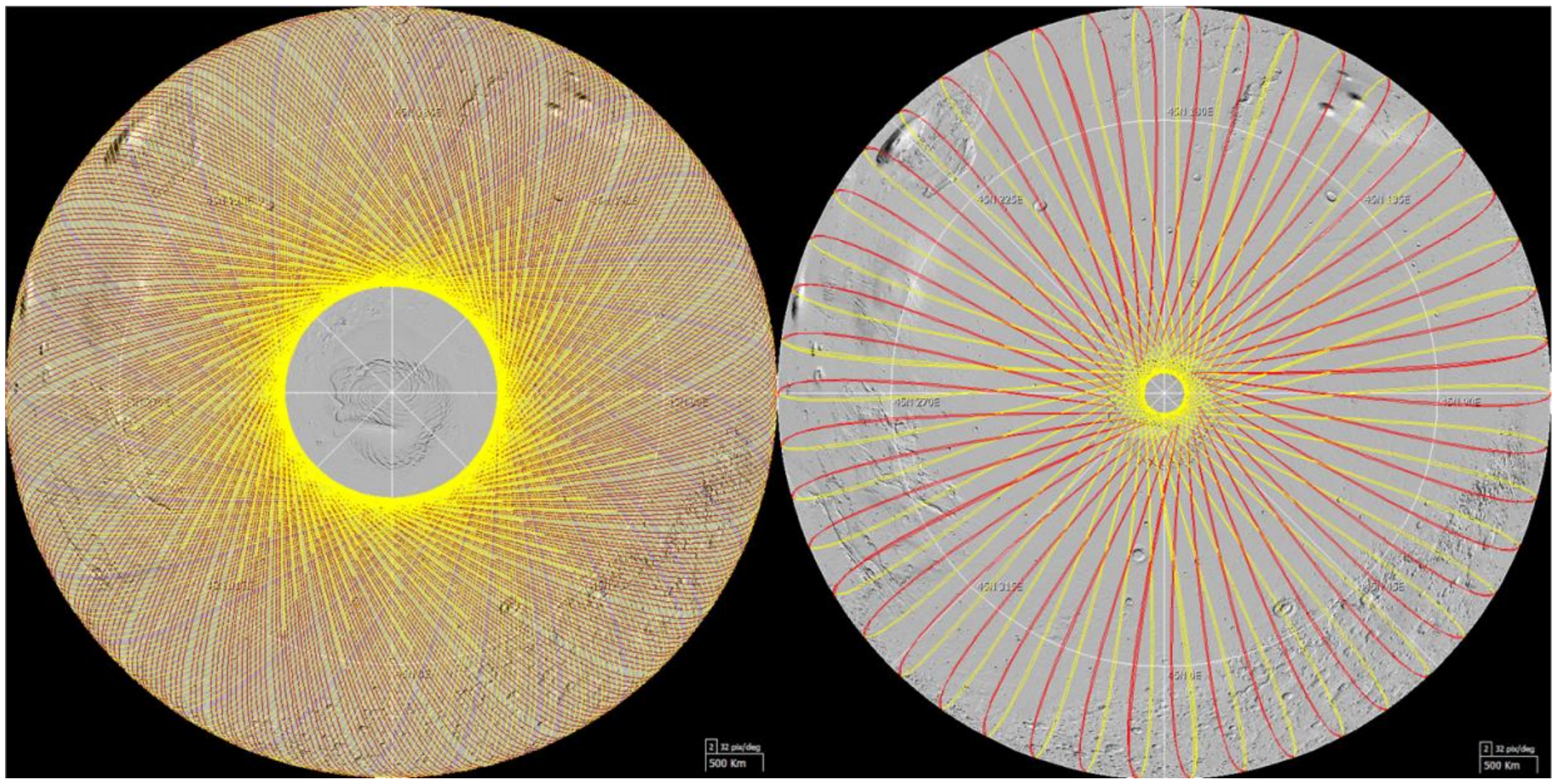

**Figure 10. Four-week (one Medium Term Planning cycle) nadir ground-tracks of TGO (left) and MEX (right) in the northern hemisphere, as seen from the North pole. Dayside tracks are shown in yellow, nightside in red. MEX has a longer orbital period, therefore the surface coverage sampling is lower than TGO's.**

When looking at the orbits of both missions in Figure 11, one can see that there are always two nodes where the ground tracks cross each other, although the distances may be very different due to the eccentricity of the MEX orbit. This does not mean that the spacecraft do actually cross these points simultaneously (they can be anywhere in the orbit with respect to each other when the orbit crossing occurs) but at least this ensures that quasi-simultaneous observations are always possible within at least 1 hour (half of the TGO orbital period). In most cases we have seen that there are always enough quasi-simultaneous opportunities with 15 minute delay or even less, depending on the season. The importance of these time differences is again highly dependent on the scientific objective of the observation. In particular, the most constraining scientific requirements are the local solar time (especially significant in the proximity of the dawn/dusk terminators) and the altitude of Mars Express based on the type of observation, as it may be interesting to observe at pericentre (for high resolution) or further towards the apocentre (for contextual information).

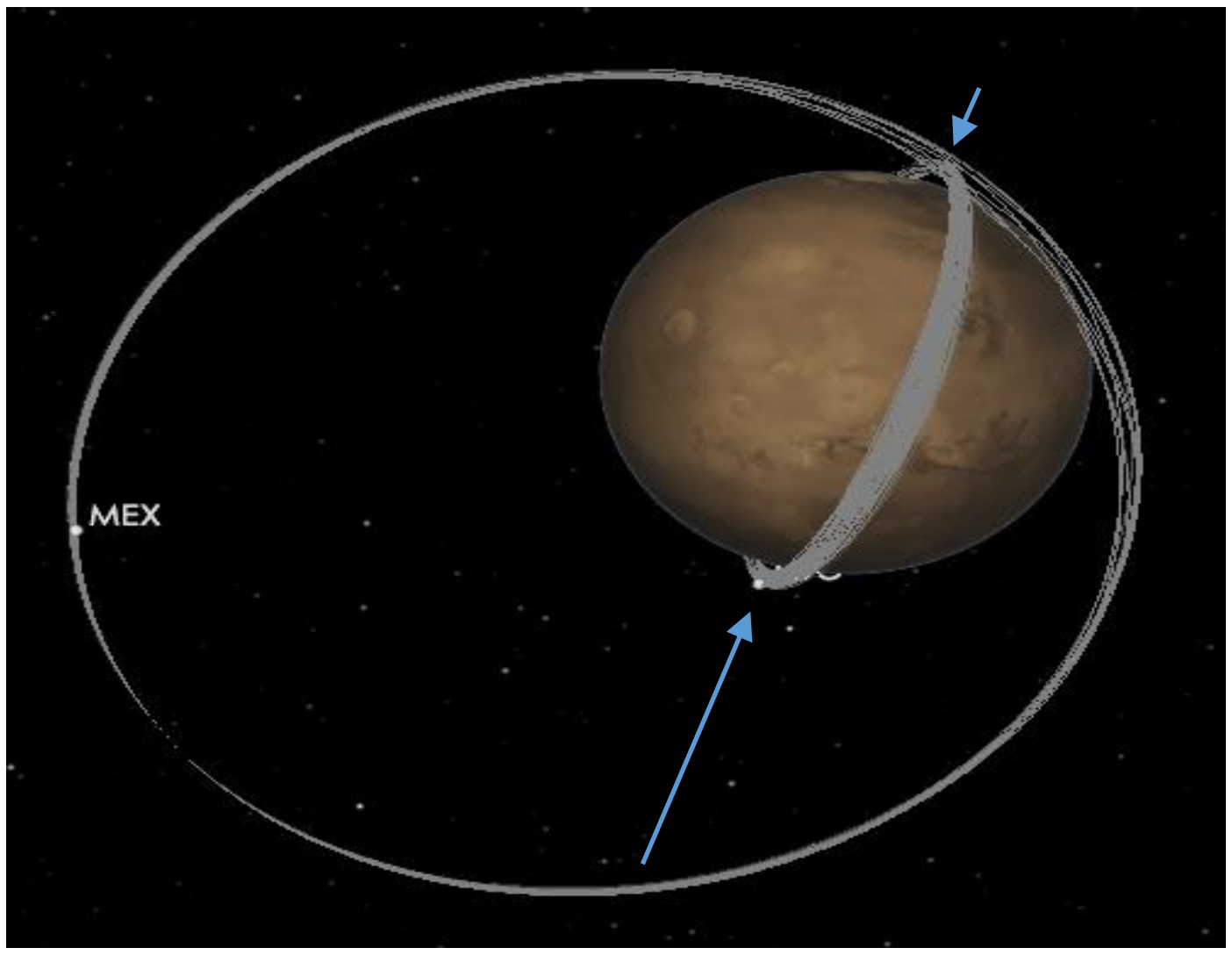

**Figure 11. 3D Simulation of MEX and TGO orbits and the crossing nodes (blue arrows) at different altitudes.**

In order to identify all the joint nadir opportunities taking into account the high variability in MEX altitudes (at pericentre and apocentre), we have performed our analysis computing the angular separation of both spacecraft as seen from Mars' centre. This angular separation, hereafter called MEX-Mars-TGO angle, is 0 degrees when both spacecraft are exactly over the same point in the surface, which would be an exact simultaneous nadir opportunity. As a first step, similarly as done for the solar occultations previously, we use a larger filter with a threshold of MEX-Mars-TGO angle < 5 degrees to identify all the possible quasi-simultaneous opportunities.

Our first analysis of combined nadir observations can be seen in Figure 12, and shows that quasi-simultaneous nadir observations are possible at different distances and illumination conditions, mostly driven by MEX pericentre seasonal evolution, and there are always two TGO-MEX orbit crossing points that can be observed by both spacecraft within less than one hour (half of the TGO orbital period). With this we were able to determine that the opportunity events occur regularly, following a latitudinal trend mostly driven by the TGO orbital node regression (~7 weeks). We can also see that the events are sparser whenever the MEX spacecraft is at pericentre, simply because the spacecraft is moving faster along the pericentre passage and therefore there are less chances of coinciding with TGO. On the other hand, when MEX is at apocentre, it moves much slower and therefore there are many more overlapping events. As previously mentioned, in most cases the pericentre observations are more scientifically interesting due to the higher resolution of the measurements, however the high-altitude observations from apocentre are also of interest as Mars Express could provide wide contextual interpretation of the atmospheric conditions to support the interpretation of nadir measurements from TGO, for example in the case of aerosols in the atmosphere (water ice, dust clouds, etc). [Määttänen, 2010; Sánchez-Lavega, 2018]

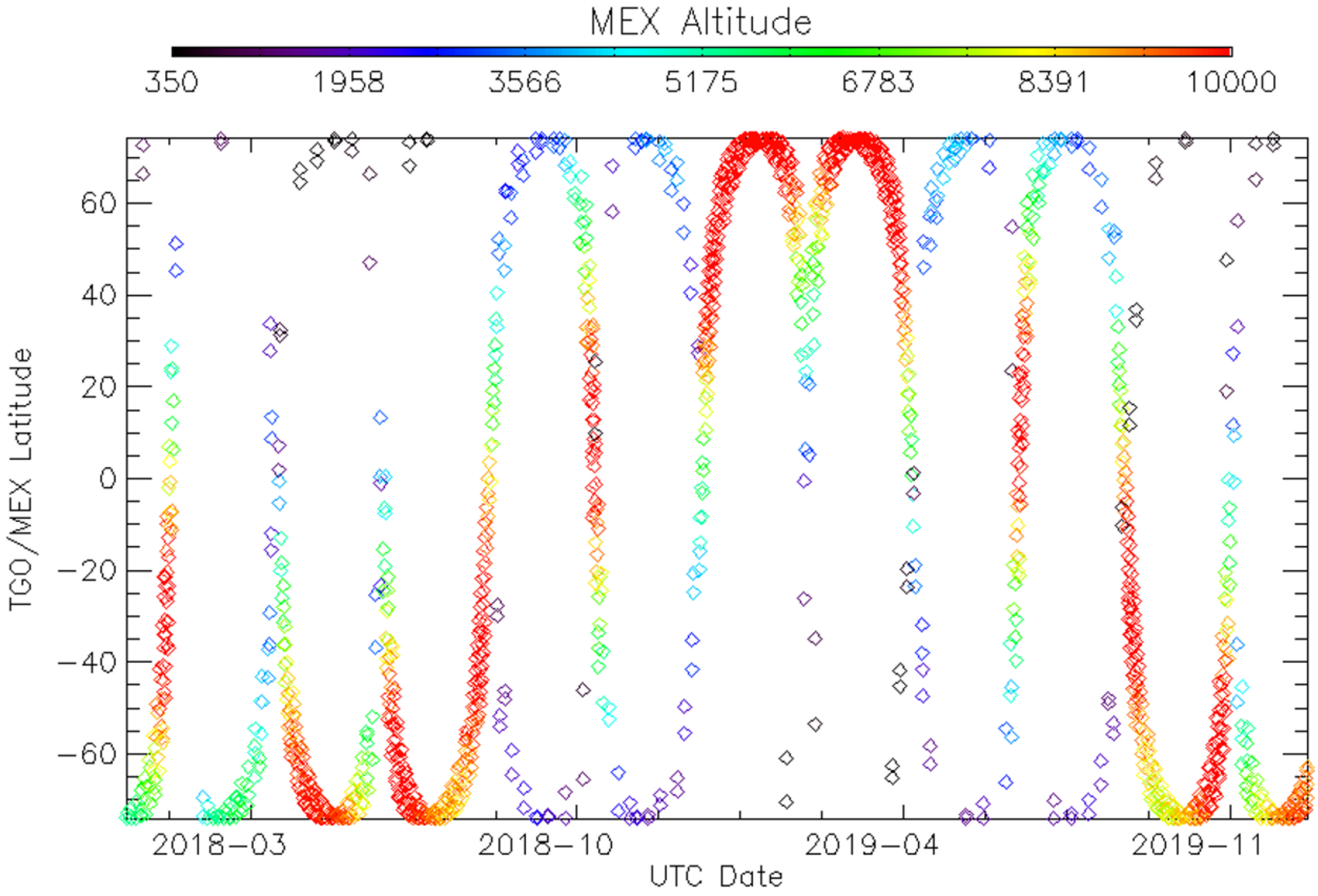

**Figure 12. MEX-TGO nadir quasi-simultaneous opportunities (whenever MEX-Mars-TGO angle < 5 deg) for the years 2018 and 2019. Note that the ground tracks cross twice per orbit, and the quasi-simultaneous opportunities are more often when the MEX altitude is higher around apocentre.**

We again obtained the long list of nadir quasi-simultaneous opportunities for the whole mission up to December 2020 and once more these were made available to the science teams of both missions. These inputs were used as a high priority during the science planning process, and joint simultaneous observations are regularly planned, typically 5~10 per month, with variable angular separations. Table 4 below provides the final list of the best simultaneous nadir observations, measured at the exact same time, that have been planned by both missions between May 2018 and May 2019. Only the simultaneous observations with a minimum angular separation are shown here (MEX-Mars-TGO angle <1.5 degrees).

| Date Time (UTC) | MEX-TGO Angle (Deg) | MEX Altitude (Km) | Solar Zenith Angle (Deg) | MEX Instruments | TGO Instruments |
|---|---|---|---|---|---|
| 2018-MAY-26 00:50:03 | *0.1* | *9917* | *64* | *SPICAM* | *NOMAD/ACS* |
| 2018-JUN-7 17:30:35 | *0.7* | *6180* | *19* | *SPICAM* | *NOMAD/ACS* |
| 2018-JUN-8 07:17:15 | *0.3* | *5601* | *14* | *SPICAM/PFS/OMEGA* | *NOMAD/ACS* |
| 2018-JUN-8 21:04:59 | *0.3* | *5002* | *7* | *SPICAM/PFS* | *NOMAD/ACS* |
| 2018-JUN-9 10:51:39 | *1.0* | *4382* | *1* | *SPICAM/PFS* | *NOMAD/ACS* |
| 2018-JUL-11 07:24:43 | *1.1* | *1042* | *88* | *PFS* | *NOMAD/ACS* |
| 2018-JUL-12 09:55:07 | *0.8* | *5171* | *93* | *PFS* | *NOMAD/ACS* |
| 2018-AUG-1 17:09:15 | *1.3* | *3116* | *95* | *PFS* | *NOMAD/ACS* |
| 2018-AUG-10 18:24:59 | *1.3* | *5021* | *61* | *PFS/OMEGA* | *NOMAD/ACS* |
| 2018-AUG-16 16:13:16 | *0.6* | *10180* | *42* | *SPICAM* | *NOMAD/ACS* |
| 2018-AUG-18 03:37:22 | *0.5* | *9478* | *45* | *SPICAM* | *NOMAD/ACS* |
| 2018-AUG-26 08:09:48 | *1.3* | *6577* | *55* | *SPICAM* | *NOMAD/ACS* |
| 2018-SEP-7 20:44:23 | *0.6* | *5248* | *58* | *PFS* | *NOMAD/ACS* |
| 2018-SEP-10 11:36:05 | *0.5* | *5194* | *58* | *SPICAM/PFS* | *NOMAD* |
| 2018-SEP-15 17:19:48 | *1.4* | *5040* | *58* | *SPICAM/PFS* | *ACS* |
| 2018-SEP-20 23:03:30 | *0.6* | *4829* | *58* | *SPICAM/PFS* | *NOMAD* |
| 2018-OCT-8 09:36:05 | *0.3* | *6735* | *67* | *SPICAM/PFS* | *NOMAD/ACS* |
| 2018-OCT-8 23:22:52 | *1.1* | *7414* | *69* | *SPICAM/PFS* | *NOMAD/ACS* |
| 2018-DEC-6 19:25:48 | *1.4* | *501* | *52* | *SPICAM/PFS* | *NOMAD/ACS* |
| 2018-DEC-9 10:18:26 | *1.2* | *422* | *49* | *SPICAM* | *NOMAD/ACS* |
| 2018-DEC-22 13:09:58 | *0.9* | *2486* | *76* | *SPICAM/PFS* | *NOMAD* |
| 2018-DEC-29 20:06:33 | *0.0* | *4493* | *91* | *SPICAM/PFS* | *NOMAD* |
| 2019-JAN-11 08:41:54 | *0.2* | *5991* | *92* | *SPICAM/PFS* | *NOMAD/ACS* |
| 2019-JAN-27 01:53:12 | *1.5* | *6460* | *83* | *SPICAM/PFS/OMEGA* | *NOMAD* |
| 2019-JAN-29 16:45:11 | *0.6* | *6541* | *81* | *SPICAM/PFS* | *NOMAD/ACS* |
| 2019-FEB-1 07:37:05 | *0.6* | *6611* | *80* | *SPICAM/PFS* | *NOMAD* |
| 2019-FEB-7 03:07:41 | *0.7* | *5999* | *70* | *SPICAM/PFS/OMEGA* | *NOMAD/ACS* |
| 2019-FEB-10 07:46:51 | *1.0* | *5287* | *61* | *SPICAM/PFS/OMEGA* | *NOMAD* |
| 2019-FEB-14 02:14:27 | *0.1* | *3685* | *41* | *SPICAM* | *NOMAD/ACS* |
| 2019-FEB-17 20:44:14 | *0.7* | *2112* | *15* | *SPICAM/PFS* | *NOMAD/ACS* |
| 2019-FEB-21 15:17:03 | *1.4* | *902* | *23* | *SPICAM* | *NOMAD/ACS* |
| 2019-FEB-27 10:51:20 | *1.2* | *569* | *45* | *SPICAM/PFS* | *NOMAD* |
| 2019-MAR-10 12:08:42 | *1.1* | *414* | *68* | *SPICAM/PFS* | *NOMAD* |
| 2019-APR-24 22:54:18 | *1.3* | *6967* | *29* | *SPICAM/PFS* | *NOMAD/ACS* |
| 2019-APR-26 10:19:54 | *1.4* | *8588* | *36* | *PFS* | *NOMAD/ACS* |
| 2019-APR-27 08:00:55 | *1.2* | *10034* | *46* | *SPICAM* | *NOMAD/ACS* |
| 2019-APR-30 20:32:27 | *0.1* | *10539* | *54* | *SPICAM* | *NOMAD/ACS* |
| 2019-MAY-2 21:40:03 | *0.9* | *10512* | *57* | *SPICAM* | *ACS* |
| 2019-MAY-6 23:55:06 | *0.8* | *10159* | *61* | *SPICAM/PFS* | *NOMAD/ACS* |
| 2019-MAY-9 01:02:32 | *0.3* | *9832* | *63* | *SPICAM/PFS* | *NOMAD/ACS* |
| 2019-MAY-13 17:01:23 | *0.3* | *9382* | *65* | *SPICAM* | *NOMAD/ACS* |

**Table 4. List of the best simultaneous nadir observations planned by TGO and MEX missions between April 2018 to June 2019. The reference times correspond to the minimum angular separation between the two spacecraft as seen from Mars' centre. The last column shows the instruments that were observing on each mission. Only the closest simultaneous observations are shown with minimum angular separation, MEX-Mars-TGO angle <1.5 degrees.**

The complete list of nadir opportunities until 2020 is provided in the supplementary information, including other various combined filtering parameters for past and future periods.

We must also note that we have only discussed here the case of nadir observations with very low MEX-TGO angle separations (<5 degrees in Figure 12, <1.5 degrees in Table 4). This implies that both spacecraft are observing almost directly down to the sub-SC ground track and the emission angle is very low. However, it is worth mentioning that some observations may also be performed at higher emission angles with some offset with respect to the ground track. This off-nadir pointing is already done routinely for Mars Express and it has already been proven feasible for Trace Gas Orbiter instruments (with some limitations, currently under discussion). These kind of off-nadir observations are out of the scope of this paper but we note that they are very useful to extend the overall coverage of the mission and they increase significantly the number of joint opportunities.

## 8. Analysis of orbit alignment periods

In addition to the joint opportunities provided in the previous sections, with detailed locations and times of the spacecraft quasi-simultaneous and simultaneous observations, another aspect to take into account is the angle between the orbit planes, which provides an indication of the observing geometry of the two spacecraft relative to each other. As illustrated in Figure 13, the relative position of MEX and TGO orbits varies from "quasi-perpendicular" (angle between orbital planes close to 90 degrees), to "quasi-aligned" (angle close to 0 degrees).

The importance of these quasi-alignment periods is not to identify specific joint observation times (already done in previous sections), but to identify seasons when the observing geometry is similar and results are therefore easily comparable.

In the case of the orbit alignment, both spacecraft fly almost in parallel, so the common sub-spacecraft regions are much larger, elongated and provide good conditions for longer joint nadir observations. Moreover during these alignment periods both orbits have similar angles with respect to the sun (beta angle) therefore the geometry of the solar occultations is also similar and it is easier to compare the results.

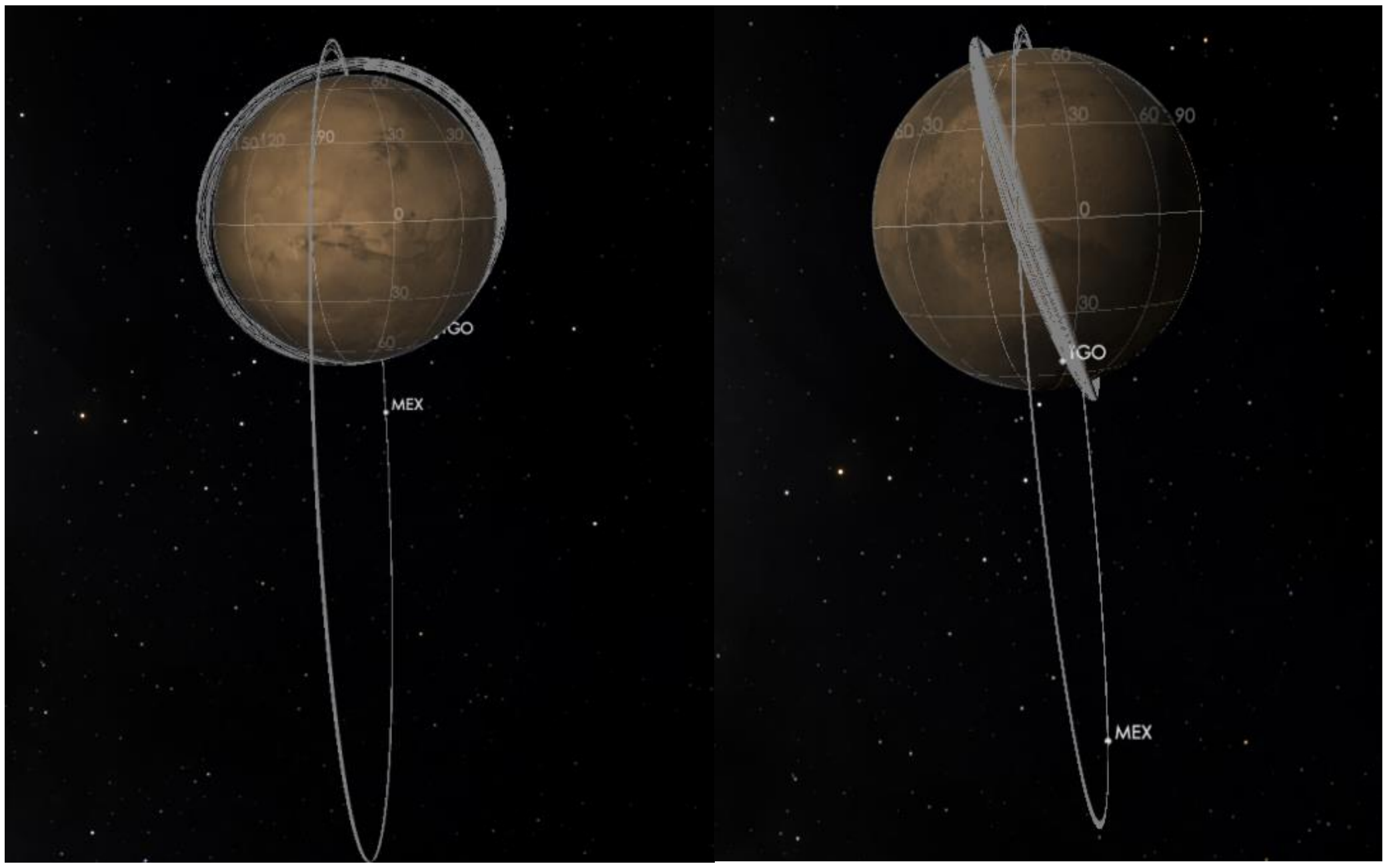

**Figure** 13. **Illustration of the variation in the angle between MEX-TGO orbital planes. "Quasi-perpendicular" case is shown on the left, "quasi-alignment" case is shown on the right.**

To identify these seasons we study the angle between MEX and TGO orbit planes, defined as the minimum angle between the two orbit planes as seen Mars equatorial plane. The evolution of the orbital angles is given in Figure 14, and is driven by the celestial mechanics affecting the TGO and MEX orbital node regression and precession.

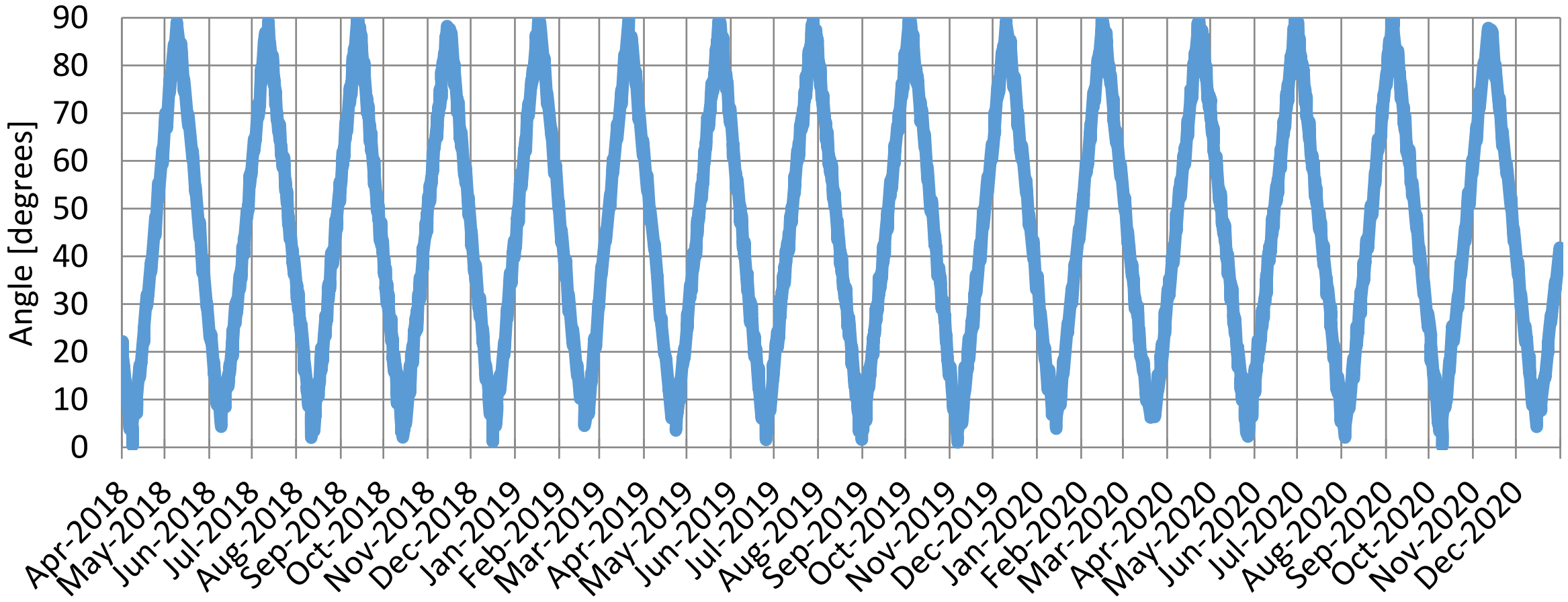

**Figure** 14. **Evolution of the angle between MEX and TGO orbit planes from April 2018 to December 2020, as measured at the Mars equatorial plane during spacecraft equator crossings. The minimum angles correspond to the MEX-TGO orbit "quasi-alignment" periods.**

These "quasi-alignment" seasons are defined as the minima of the angle between the orbit planes. We provide in Table 5 the central dates corresponding to these periods, which occur regularly every 9~10 weeks. For each of these periods we provide the phase angle observed at both equator crossings (ascending and descending nodes). This phase angle is a direct indication of the geometry with respect to the sun, which can be used directly to recognize the main nadir illumination and solar occultation seasons. For example, when the spacecraft fly over the terminator (phase angle 85~95º) the nadir observations are badly illuminated and no sun occultation are possible. On the other hand, when the spacecraft fly closer to the sub-solar and anti-solar points (phase angle 0º and 180º respectively), they allow for well illuminated dayside nadir observations and good vertical profiles with solar occultations.

We note in Table 5 also the relative orientation of the flight directions, which alternate for each period. That is, in some quasi-alignment periods both spacecraft fly in the same direction (i.e. the ascending nodes with both spacecraft flying south to north are on the same side), whereas in the next period the spacecraft will fly in opposite directions (one spacecraft crossing north to south, while the other is flying south to north).

| ***Periods of Orbit Quasi-Alignment*** | ***Phase Angle (deg) at equator crossing*** | | ***Flight Direction*** |
|---|---|---|---|
| ***Central Date*** | ***Min*** | ***Max*** | ***Orientation*** |
| *08/04/2018* | *35* | *145* | *Same* |
| *10/06/2018* | *0* | *180* | *Opposite* |
| *11/08/2018* | *45* | *135* | *Same* |
| *15/10/2018* | *85* | *95* | *Opposite* |
| *16/12/2018* | *45* | *135* | *Same* |
| *18/02/2019* | *10* | *170* | *Opposite* |
| *21/04/2019* | *25* | *155* | *Same* |
| *24/06/2019* | *60* | *120* | *Opposite* |
| *31/08/2019* | *85* | *95* | *Same* |
| *06/11/2019* | *55* | *125* | *Opposite* |
| *14/01/2020* | *25* | *155* | *Same* |
| *23/03/2020* | *25* | *155* | *Opposite* |
| *27/05/2020* | *60* | *120* | *Same* |
| *03/08/2020* | *70* | *110* | *Opposite* |
| *10/10/2020* | *25* | *155* | *Same* |
| *15/12/2020* | *25* | *155* | *Opposite* |

**Table 5. List of the TGO and MEX quasi-alignment periods (minimum angle of MEX TGO orbits as seen at the equator), between April 2018 and December 2020. The sun phase angles are given as seen at the equator crossing points. Note that, as the equator is crossed twice in the orbit, there are always two complementary values (e.g. 0º/180º). The last column represents the relative orientation of the flight directions, as both spacecraft may be flying in the same or in opposite directions.**

## 9. Analysis of future coordinated observations

### 9.1. MEX Nadir / TGO Solar Occultations

Based on the successful observations in nadir and solar occultation modes that were planned in 2018, it has been discussed within the frame of the TGO and MEX Science Working Teams to investigate the possibility to use more advanced observation geometries, in particular to observe the TGO Sun Occultation tangent point coinciding with Mars Express sub-satellite point(s), as illustrated in Figure 15. This mode of observation has been requested by various instrument teams in order to measure a vertical profile of the atmosphere with NOMAD and/or ACS instruments at the solar occultation point, while obtaining the atmospheric and surface temperature of the MEX sub-spacecraft point with the PFS instrument in nadir mode. Combination of the limb and nadir geometries used by two spacecraft would enable better precision and larger vertical range of temperature sounding as well as putting additional constraints on the vertical distribution of minor species. Other observations may also be interesting to perform with other instruments in the future, for example to monitor larger atmospheric features such as water or $CO_2$ ice clouds at the terminator captured by the cameras (OMEGA, HRSC, and VMC) [Määttänen, 2010; Sánchez-Lavega, 2018].

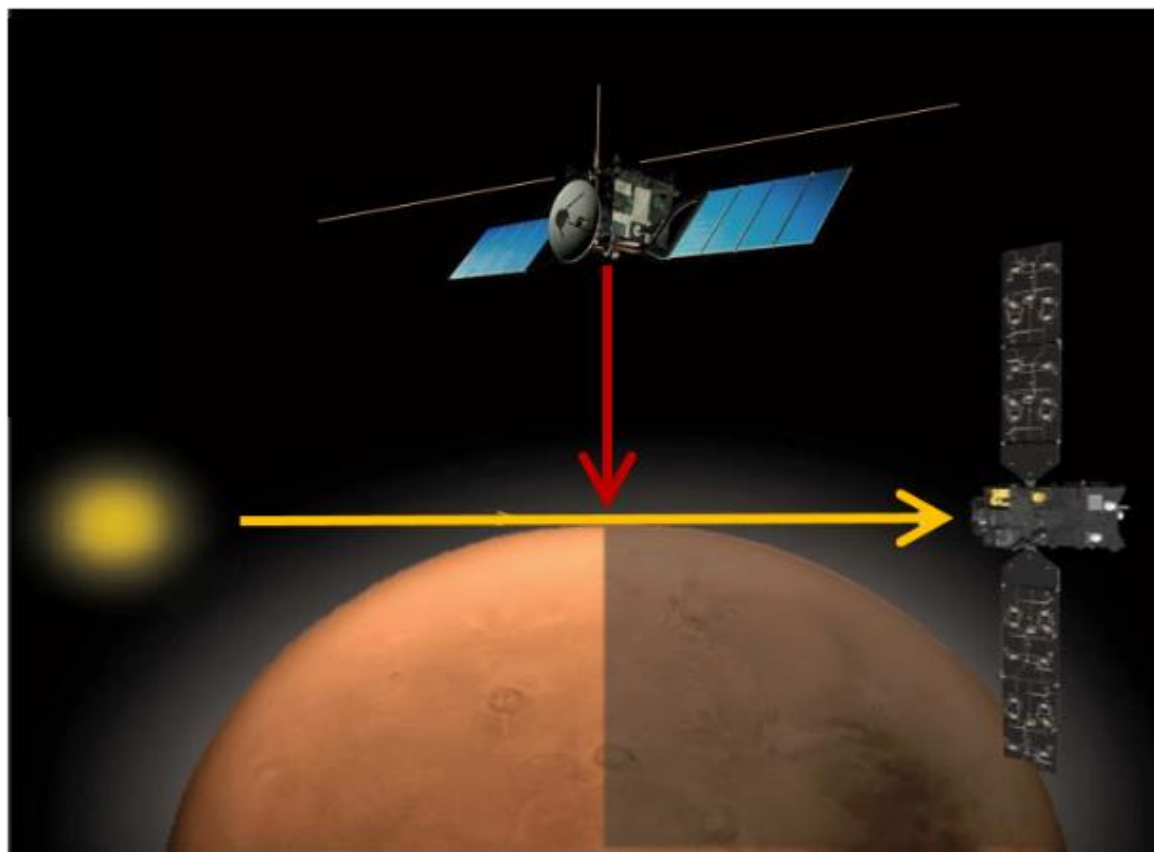

**Figure 15. Illustration of MEX nadir observation of the TGO solar occultation tangent point in the atmosphere.**

The computations of the combined nadir / solar occultation points have been performed for the whole of 2019, showing that indeed these opportunities occur regularly, mostly driven by TGO occultation seasons, and coinciding with a spatial difference down to a few degrees in latitude and longitude.

Table 6 below provides the full list of simultaneous observations for 2019 and 2020 within a latitude and longitude difference of 10 degrees between the MEX sub-spacecraft point and the TGO solar occultation tangent point. The complete list of opportunities for 2019 and 2020 with wider filters is provided in the supplementary information. We note again that the solar occultation measurements cover a wide range of spatial coordinates along the altitude profile, highly dependent on the observing geometry. However, for the scope of this long term opportunity analysis it has been agreed to use the 0 km reference altitude for simplicity since it is sufficient to identify the best opportunities using wide latitude/longitude filters.

| Date Time (UTC) | TGO Sun Occultation | | | MEX Nadir | | | Differences | | |
|---|---|---|---|---|---|---|---|---|---|
| | Latitude (deg) | Longitude (deg) | Beta Angle (deg) | Latitude (deg) | Longitude (deg) | Altitude (km) | Latitude (deg) | Longitude (deg) | Distance (km) |
| 2019-02-10T20:30:49 | -77 | 321 | 34 | -69 | 316 | 621 | 8 | 5 | 509 |
| 2019-03-06T20:42:25 | 86 | 71 | 44 | 87 | 75 | 10545 | **1** | **4** | **44** |
| 2019-04-08T12:38:13 | 82 | 204 | 22 | 74 | 209 | 7987 | 9 | 5 | 512 |
| 2019-04-09T02:23:12 | 82 | 358 | 20 | 79 | 3 | 8587 | **3** | 5 | **169** |
| 2019-04-09T16:08:15 | 81 | 152 | 18 | 84 | 149 | 9150 | **3** | **3** | **170** |
| 2019-04-12T06:59:58 | 77 | 301 | 10 | 85 | 301 | 9150 | 8 | **0** | 467 |
| 2019-04-24T00:36:51 | -82 | 318 | 27 | -86 | 315 | 518 | **4** | **3** | 236 |
| 2019-04-26T15:28:37 | -83 | 132 | 34 | -85 | 126 | 560 | **1** | 6 | **95** |
| 2019-06-22T01:34:59 | 53 | 335 | 49 | 57 | 334 | 2314 | **4** | **0** | 254 |
| 2019-07-17T17:45:29 | -49 | 147 | 53 | -40 | 146 | 5548 | 9 | **0** | 535 |
| 2019-07-18T07:31:30 | -51 | 307 | 51 | -49 | 304 | 4616 | **2** | **3** | **151** |
| 2019-07-18T21:17:25 | -52 | 108 | 49 | -60 | 101 | 3640 | 8 | 7 | 529 |
| 2019-09-03T06:02:36 | -18 | 43 | 49 | -19 | 33 | 10430 | **1** | 10 | 537 |
| 2019-09-03T19:48:53 | -15 | 200 | 48 | -23 | 192 | 10279 | 8 | 8 | 651 |
| 2019-09-04T03:41:07 | -13 | 85 | 48 | -8 | 78 | 10379 | 5 | 6 | 477 |
| 2019-09-04T17:27:18 | -11 | 242 | 48 | -12 | 236 | 10462 | **1** | 6 | 344 |
| 2019-09-05T07:13:36 | -8 | 40 | 47 | -15 | 34 | 10492 | **7** | 5 | 535 |
| 2019-09-05T15:05:48 | -6 | 284 | 46 | 0 | 281 | 9858 | 6 | **3** | 385 |
| 2019-09-06T04:52:02 | -3 | 82 | 46 | -4 | 79 | 10074 | **1** | **3** | **182** |
| 2019-09-06T18:38:11 | 0 | 239 | 45 | -8 | 237 | 10300 | 8 | **2** | 477 |
| 2019-09-07T02:30:28 | 2 | 124 | 44 | 10 | 123 | 8795 | 8 | **1** | 484 |
| 2019-09-07T16:16:35 | 4 | 281 | 43 | 4 | 282 | 9321 | **0** | **0** | **24** |
| 2019-09-08T06:02:43 | 7 | 79 | 42 | 1 | 79 | 9605 | 6 | **1** | 354 |
| 2019-09-09T03:41:02 | 11 | 121 | 41 | 16 | 124 | 7980 | **4** | **3** | 299 |
| 2019-09-09T17:27:03 | 14 | 279 | 40 | 10 | 282 | 8566 | **4** | **4** | 295 |
| 2019-09-11T04:51:13 | 20 | 119 | 37 | 22 | 125 | 7010 | **2** | **7** | 394 |
| 2019-09-11T18:37:07 | 22 | 276 | 35 | 16 | 284 | 7735 | 6 | **7** | 555 |
| 2019-10-15T07:56:21 | -48 | 207 | 57 | -55 | 211 | 2794 | 7 | **4** | 432 |
| 2019-11-09T14:20:37 | 48 | 163 | 45 | 39 | 161 | 2347 | 9 | **2** | 544 |
| 2020-05-07T02:42:44 | -75 | 157 | 6 | -71 | 148 | 2578 | **3** | 8 | 245 |
| 2020-05-09T17:36:43 | -70 | 311 | 2 | -68 | 310 | 2290 | **2** | **1** | **133** |
| 2020-05-11T19:26:23 | 77 | 112 | 9 | 82 | 107 | 2056 | 5 | 5 | 298 |
| 2020-05-12T08:30:54 | -65 | 107 | 10 | -63 | 112 | 1933 | **3** | 5 | **199** |
| 2020-05-14T10:20:10 | 79 | 281 | 17 | 79 | 274 | 2316 | **0** | 7 | **81** |
| 2020-05-14T23:25:23 | -59 | 265 | 18 | -56 | 273 | 1585 | **3** | 8 | 313 |
| 2020-06-06T10:48:29 | 52 | 123 | 60 | 46 | 127 | 6354 | 6 | 5 | 411 |
| 2020-06-07T00:35:47 | 46 | 279 | 60 | 53 | 285 | 5559 | 7 | 6 | 480 |
| 2020-08-06T17:55:56 | -23 | 226 | 58 | -25 | 219 | 4849 | **2** | 6 | 364 |
| 2020-08-08T18:59:28 | -39 | 219 | 52 | -34 | 222 | 3915 | 5 | **3** | 322 |
| 2020-10-05T05:08:39 | 64 | 66 | 23 | 73 | 58 | 7556 | 9 | 8 | 588 |
| 2020-10-07T20:02:35 | 65 | 228 | 16 | 74 | 218 | 7804 | 9 | 10 | 575 |
| 2020-12-24T01:06:21 | 83 | 117 | 43 | 74 | 121 | 10244 | 9 | **3** | 562 |
| 2020-12-25T12:28:38 | 84 | 336 | 47 | 81 | 332 | 9645 | **3** | **4** | **193** |

**Table 6. List of combined opportunities for MEX nadir and TGO solar occultation points in 2019 and 2020. The time and solar occultation coordinates correspond to the crossing at 0 km altitude. The MEX spatial coordinates are computed at the time of the occultation. The differences are filtered by latitude and longitude below 10 degrees.**

## 9.2. Future coordinated limb observations

We have so far described the two main TGO observation modes: nadir for global atmospheric monitoring and solar occultation for detailed vertical atmospheric profiles. Unfortunately, these solar occultations can only provide profiles limited to the terminator region (dusk and dawn) and cannot cover other local times, in particular the sub-solar and anti-solar regions (noon, midnight) cannot be covered. In order to overcome this limitation [Lopez-Valverde, 2018] proposed an off-the-terminator limb pointing. This allows observation of various scientifically interesting emissions in the infrared, visible and ultraviolet such as nightglows (aurorae, $O_2$, NO, O, OH, …), dayglow and fluorescent emissions, non-Local Thermodynamical Equilibrium (non-LTE) emissions, CO and $CO_2$, high-altitude clouds, etc. as described in [Lopez-Valverde, 2018], thus extending atmospheric investigations to a broad range of local times.

Limb observations can be performed by TGO in two ways. First simply observing with the solar boresight, fixed with respect to the spacecraft, which points at the limb even when the Sun is not in the field of view. This can be done only with two channels, NOMAD's LNO and ACS' TIRVIM, and is already being planned routinely via commanding at STP level, without any further implications to the pointing. The second possibility is to rotate the whole spacecraft to point the nadir boresight to the limb, allowing observations with UVIS, TIRVIM and NIR nadir channels, which have more sensitivity since they are designed to look directly at Mars atmosphere. This spacecraft rotation has already been tested and is now routinely planned, typically once a week with dedicated inertial pointings towards the sub-solar point.

In summary, TGO is now capable of observing the limb routinely so here we present the first analysis of joint limbs than can be observed simultaneously with MEX. These possible joint observations will open a new field of collaboration for various scientific areas. In particular, for this study we will focus on dayside limbs, as there is a high interest in comparing the non-LTE emission of $CO_2$ at 4.3µm with the MEX/OMEGA infrared long channel and the same emission with TGO/ACS TIRVIM, or TGO/NOMAD LNO at 2.7 µm.

As a starting point we have analysed the overall limb coverage that can be obtained by TGO itself. In Figure 16 we can see the latitude, longitude and illumination (given as sun elevation angle, complementary of the solar incidence angle) of the limb tangent points in the atmosphere as seen from the TGO spacecraft towards the sub-solar point, that is the brightest limb possible in the horizon. Figure 16 demonstrates that TGO can cover the whole dayside at various illuminations (sun elevation >10 degrees to avoid the terminator region where solar occultations occur). The evolution of the coverage is driven by the beta angle, as shown in Figure **17**.

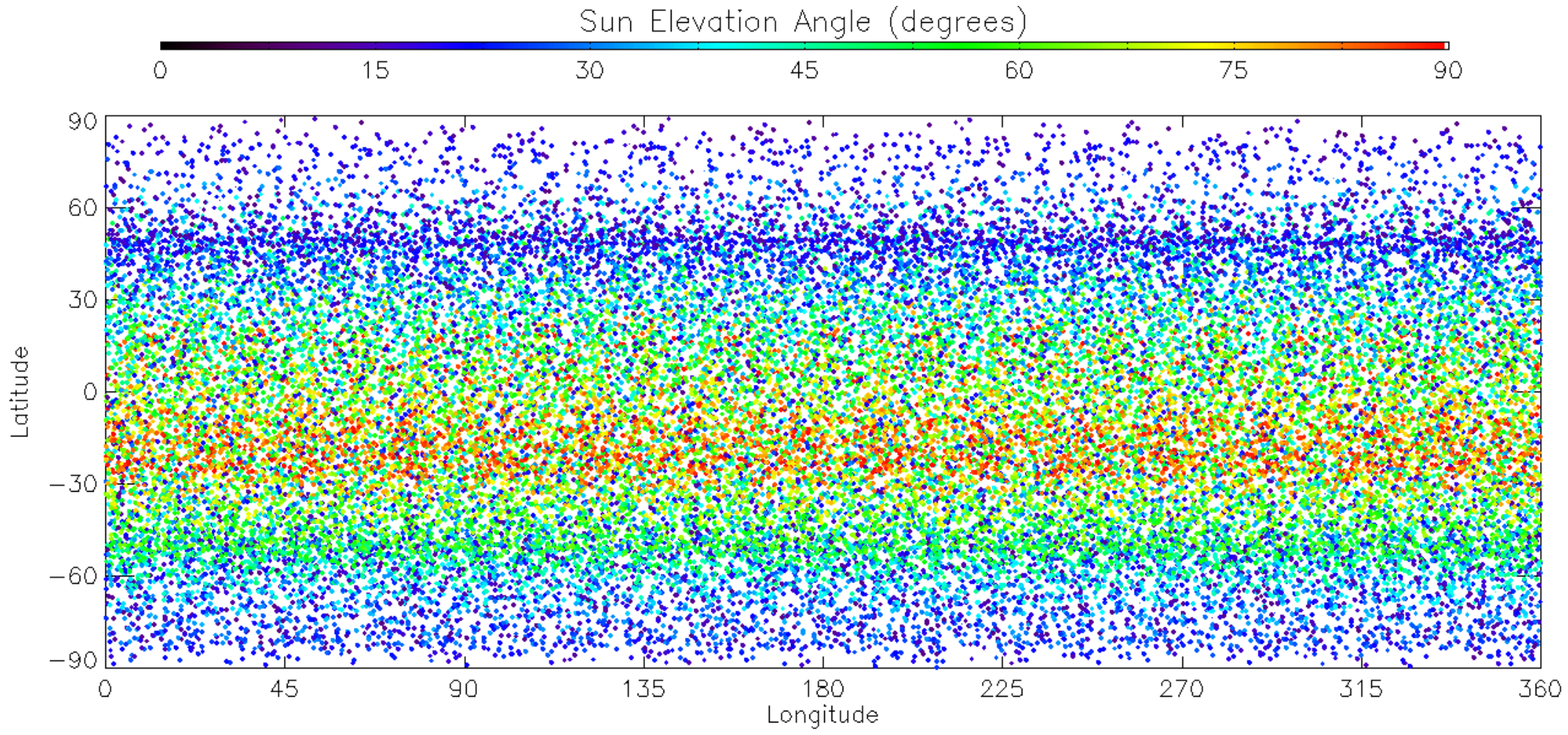

**Figure 16. Coverage of the brightest limb tangent point in the atmosphere as seen by TGO. Latitude and longitude spatial coverage for the whole 2020, with a sampling of 10 minutes. Coloured by solar elevation angle. (black is the minimum, limited to 10 degrees to avoid terminator; red is the maximum, 90 degrees at the sub-solar point)**

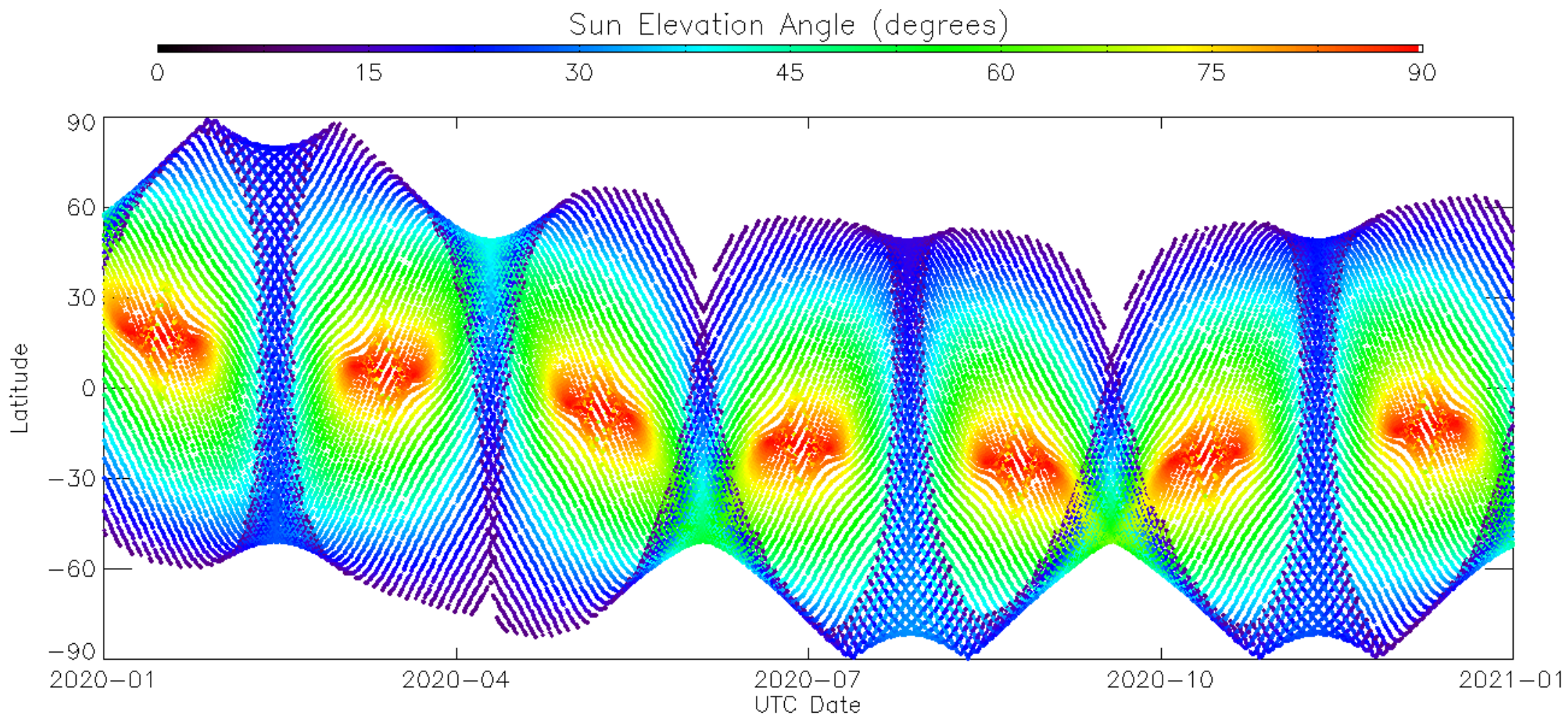

**Figure 17. Coverage of the brightest limb tangent point in the atmosphere as seen by TGO. Latitudinal variation over time for 2020, with a sampling of 10 minutes. Coloured by solar elevation angle. (black is the minimum, limited to 10 degrees to avoid terminator; red is the maximum, 90 degrees at the sub-solar point)**

We now present the analysis of joint limb opportunities between MEX and TGO. As previously mentioned, MEX has more flexible pointing constraints and therefore it can virtually point anywhere across the horizon line. In order to simplify our analysis, we have computed only three possible limb cases. Two limbs are computed in the orbital plane in the forward and backward direction, providing a vertical profile of the atmosphere in front of or behind the spacecraft. The third limb is computed at the brightest illumination possible, that is the closest to the sub-solar point as this is one of the most interesting regions.

The results of the opportunity analysis for the simultaneous opportunities between both missions are shown in**Error! Reference source not found.** Figure 18 and Figure 19. These figures indicate only the opportunities that could be possible and not the actual observations planned, since this is only proposed as a possibility to be implemented in the future.

Here the full list of filtering parameters used for the analysis:

- TGO limb at the brightest point in the horizon, closest to sub-solar point, at a reference altitude of 50 km
- MEX limb as seen at the forward, backward and brightest points, at a reference altitude of 50 km
- Dayside limbs with solar elevation angle >1 degree (excluding pure terminator and solar occultations)
- Difference in latitude and longitude <10 degrees
- Simultaneous observations within <1 min time difference
- Sun elevation angle >10 degrees to exclude terminator where the solar occultations occur.
- MEX distance to the limb <10,000 km, approximately corresponding to <7,000 km altitude of the MEX spacecraft, or +/-90 minutes around MEX pericentre.

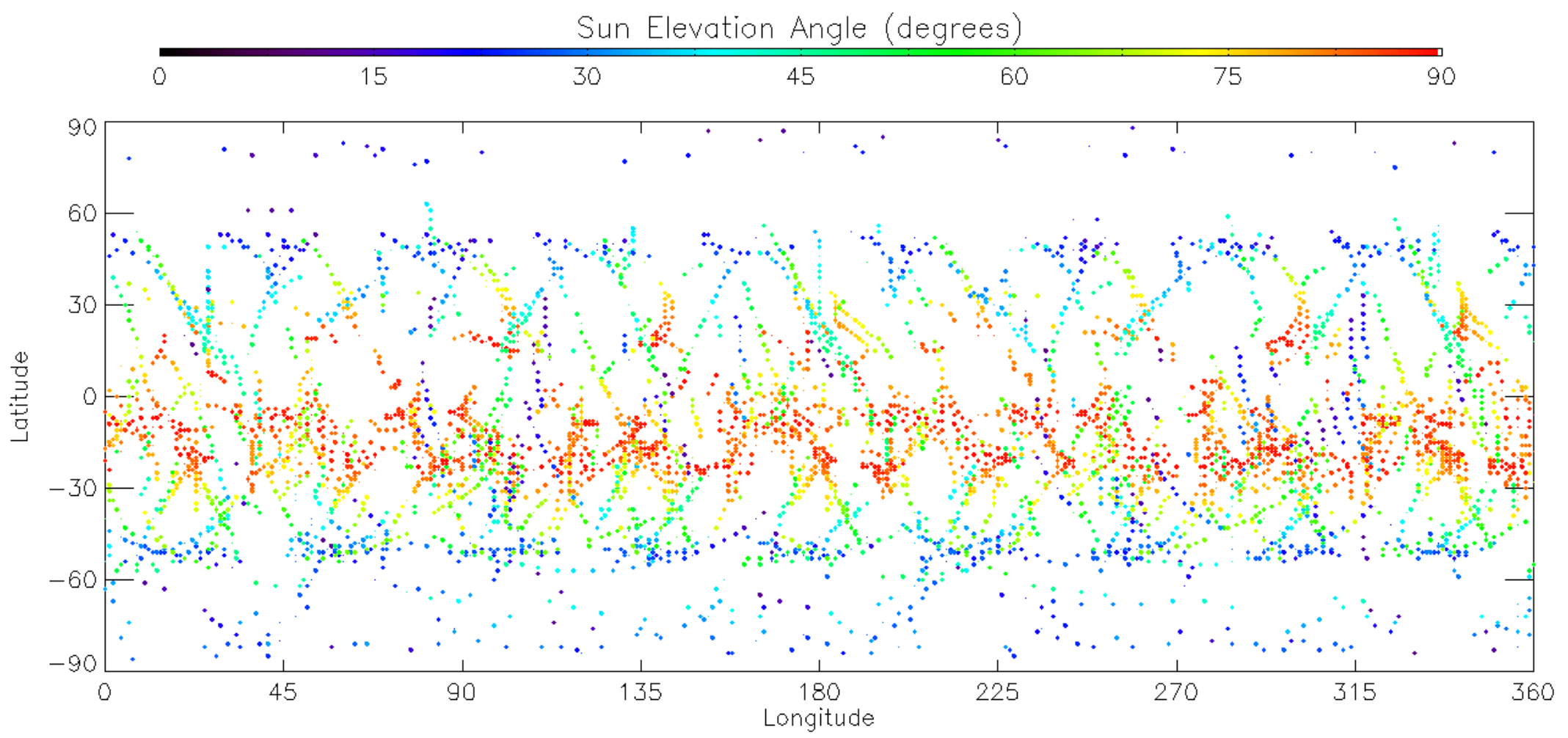

**Figure 18. Spatial coverage (latitude vs longitude) of the simultaneous limb opportunities by MEX and TGO for 2019. Coloured by solar elevation angle. (black is the minimum, limited to 10 degrees to avoid terminator; red is the maximum, 90 degrees at the sub-solar point)**

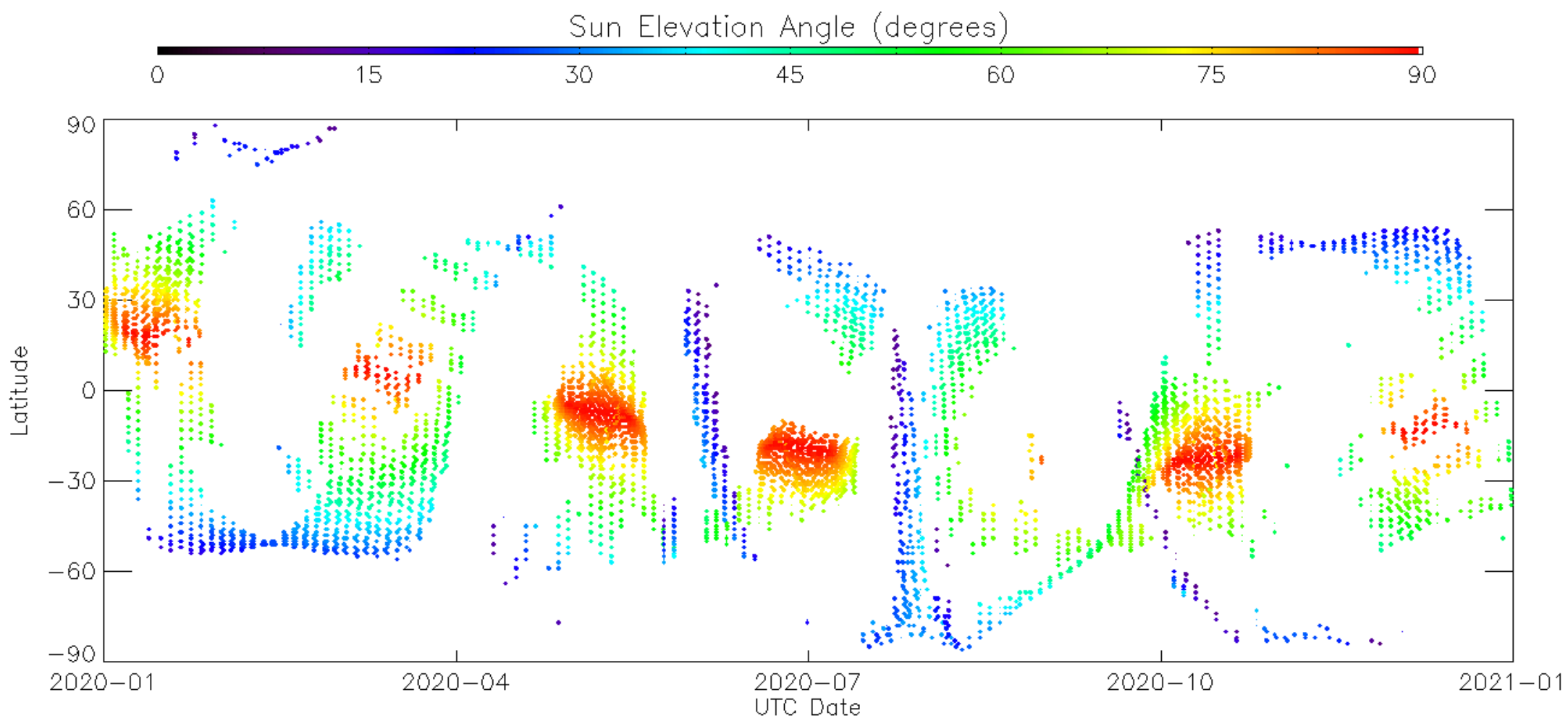

**Figure 19. Latitude coverage over time of the simultaneous limb opportunities by MEX and TGO for 2019. Coloured by solar elevation angle. (black is the minimum, limited to 10 degrees to avoid terminator; red is the maximum, 90 degrees at the sub-solar point)**

This science opportunity analysis confirms that the simultaneous co-located limbs can be performed regularly and cover a wide range of latitudes, longitudes and illumination angles. These opportunities reach the sub-solar region fulfilling the scientific requirement to observe non-LTE emissions and various other dayglows mentioned by [Lopez-Valverde, 2018].

In summary, with this study we confirm that combined limb observations by MEX and TGO are possible. The first test observations will be planned in 2020 and we expect to implement them routinely throughout the next years. We note that TGO limb observations are performed during the routine nadir in between solar occultations, and so they may have an impact in the nadir coverage. For this reason, their usage is limited and needs to be discussed in the frame of the science working team to find the right balance matching the scientific objectives of the mission.

### 9.3. Future coordinated spacecraft-to-spacecraft radio occultations

Scientific return of both Mars Express and Trace Gas Orbiter missions can be significantly enhanced by performing spacecraft-to-spacecraft radio occultations using the UHF relay system that is designed to communicate with the various surface rovers. This investigation, whose technical feasibility is currently under study, would be very useful to retrieve ionospheric and atmospheric profiles at Mars. Radio waves propagating through a refractive medium like the Martian atmosphere or ionosphere experience phase shifts, which can be used to study the refraction properties of the medium along the signal ray path. Converting these refractivity profiles into vertical profiles of electron density, neutral number density, temperature and pressure can give important information about structure and dynamic processes of the ionosphere and lower/middle atmosphere.

The spacecraft-to-spacecraft radio occultations performed in UHF are primarily sensitive to the ionosphere and can significantly extend spatial, local time and seasonal coverage of current spacecraft-to-Earth measurements, since they are not limited to the Earth occultation seasons, driven by planetary ephemeris, and they are not restricted by the availability of ground stations.

For this purpose, we have expanded our scientific opportunity analysis to compute the MEX-Mars-TGO radio occultation points. These points are defined in the same way as the solar occultation previously mentioned, the intercept between the spacecraft to spacecraft vector with the Mars reference ellipsoid (equatorial and polar radius 3389.50 and 3396.19 as per [Archinal, 2010]), at a limb height of 0 km with respect to this ellipsoid.

We note again that the radio occultation measurements would cover a wide range of spatial coordinates along the altitude profile, which would depend highly in the observing geometry, and affected by the radio frequency and signal to noise ratio. However, for the scope of this preliminary analysis it has been agreed to use the 0 km reference altitude since it is sufficient to identify the overall coverage capabilities.

The plots in Figure 20 show the evolution of the MEX-Mars-TGO radio occultation points' latitudinal coverage throughout the year 2020 and the total spatial coverage (latitude, longitude) over the same year. This illustrates the excellent spatial and temporal coverage of the atmosphere that could be achieved in this configuration. The distance (MEX to TGO) is also shown as colours in Figure 20, providing a first indication of the potential variability in the signal to noise ratio that could be achieved with the radio link. The first demonstration of such mutual occultation measurements at Mars were performed with NASA's Mars Odyssey and Mars Reconnaissance Orbiter spacecraft [Ao et al., 2015].

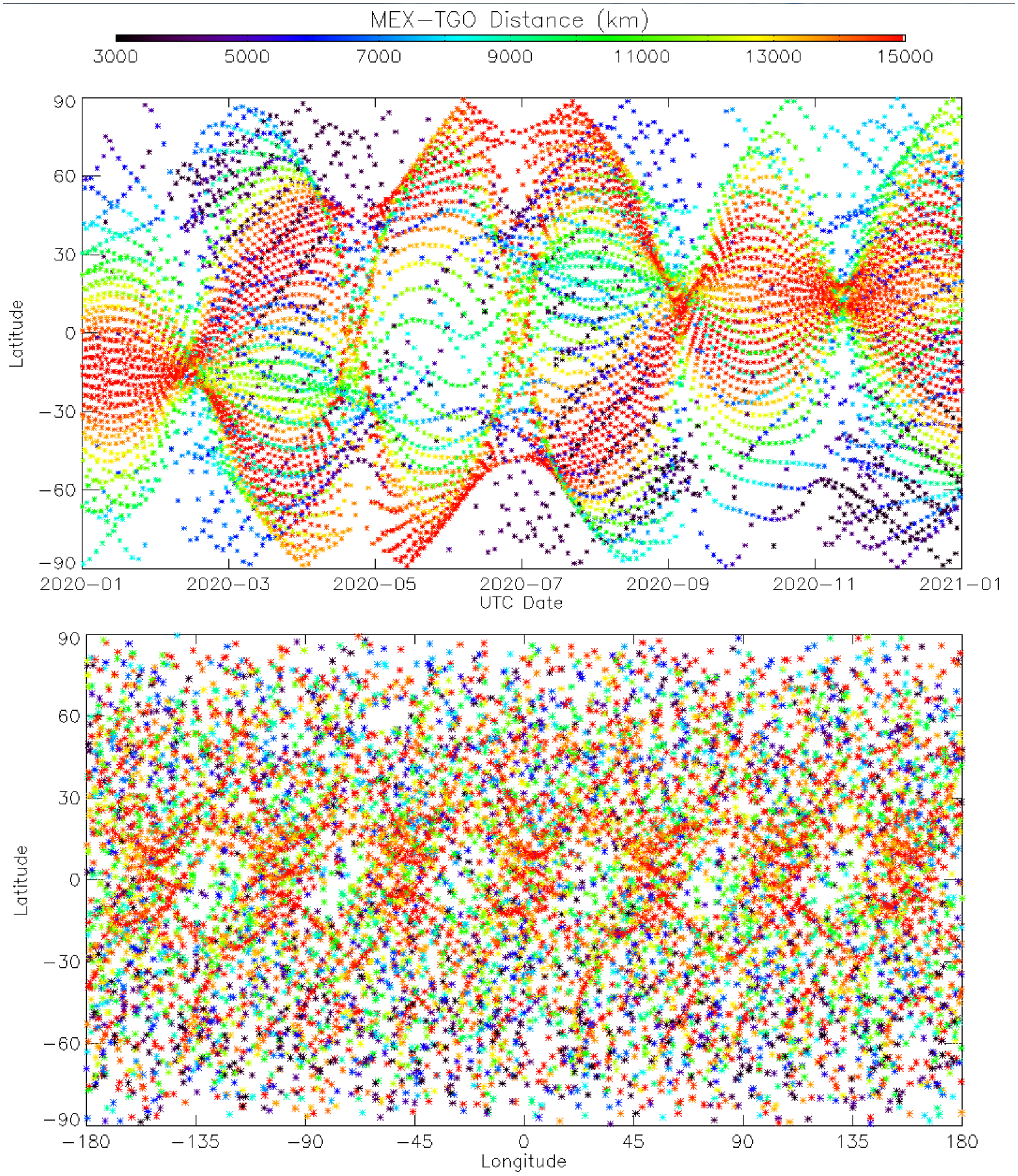

**Figure 20. Simulation of possible spacecraft-to-spacecraft occultation opportunities in the first half of 2019. The MEX-Mars-TGO occultation points are shown in latitudinal coverage through time (top) and spatial latitude/longitude coverage (bottom). The colour represents the distance between both spacecraft.**

This study the first step of a large dedicated feasibility analysis currently being done within ESA, including a full study of the spacecraft capabilities, radio signal propagation through the atmosphere and technical configuration of the radio antennas in collaboration with NASA. The first operational tests could be implemented in 2020 and the final implementation is expected to take place in 2021, as part of the next mission extension phase for both Mars Express and Trace Gas Orbiter.

## 10. Conclusions

In this paper we have identified and analysed the joint science observations that have been coordinated between the Mars Express and Trace Gas Orbiter satellites since the start of the nominal TGO science phase in April 2018 until June 2019, and the future potential opportunities for further combined observations until December 2020.

We have identified the combined solar occultation opportunities, with many quasi-simultaneous observations within a 15-minute difference that have already been planned. We have also identified and planned simultaneous observations within a difference of less than 2 minutes, which can be used for cross-calibration of the instruments and comparison of vertical atmospheric profiles. All latitudes and local times can be observed by both missions and the opportunities are mainly driven by the limited duration of the MEX solar occultation seasons.

We have also analysed the combined nadir opportunities that can be exploited by both missions. The simultaneous cross-calibrations are possible regularly at different distances and illumination conditions. There are always two TGO-MEX orbit crossing points that can be observed within one-hour difference due to shorter TGO orbital period. All latitudes and longitudes can be covered by both spacecraft, with the exception of the polar regions due to the TGO orbit inclination, and the opportunities pattern is mainly defined by the MEX pericentre seasonal variation. Also there are seasons with orbit "quasi-alignment", occurring every few months, when both spacecraft fly "quasi-parallel" to each other, with extended overlapping of the nadir ground tracks on the surface.

Finally, we have performed an analysis of future combined opportunities, including the possible combination of MEX nadir and TGO solar occultation points, combined limb observations and a preliminary analysis for future spacecraft-to-spacecraft radio occultations that demonstrates a very good spatial and temporal coverage of the atmosphere.

The opportunities identified in this study are routinely provided as an input for both Mars Express and ExoMars TGO science planning teams. The resulting observations strongly increase robustness of both Mars Express and Trace Gas Orbiter investigations due to cross-calibration of the instruments, significantly enhance spatial and temporal coverage and open new opportunities for future scientific collaborations and synergies within the scientific community working for both missions.

We note that beyond the coordination efforts described in this paper there are also multiple initiatives for scientific and operational collaboration with other Mars missions by NASA (MSL, MRO, MAVEN, etc) and other agencies (India, Japan, China, …). These interactions are beyond the scope of this study, but we can confirm that they are already in place at various levels, with dedicated working groups sharing results and preparing ad-hoc observation campaigns (e.g. methane search, atmospheric escape, dust monitoring, etc).

## 11. Acknowledgements

The authors acknowledge the contributions of the European Space Agency, Roscosmos, the *Centre National d'Études Spatiales* (CNES) and all other National Agencies, research institutions and teams involved in the success of the Mars Express and ExoMars 2016 mission. Special thanks to M. Giuranna, E. Hauber, A.C. Vandaele, I. Thomas, F. Daerden, N. Ignatiev, A. Fedorova, A. Trokhimovsky, O. Korablev, and in general to all the instrument teams that have supported this work. We also thank the valuable contributions given by two anonymous reviewers to improve this manuscript.

ExoMars is a space programme of the European Space Agency (ESA) and Roscosmos. The NOMAD experiment is led by the Royal Belgian Institute for Space Aeronomy (IASB-BIRA), assisted by Co-PI teams from Spain (IAA-CSIC), Italy (INAF-IAPS), and the United Kingdom (Open University). This project acknowledges funding by the Belgian Science Policy Office (BELSPO), with the financial and contractual coordination by the ESA Prodex Office (PEA 4000103401, 4000121493), by the Spanish Ministry of Science and Innovation (MCIU) and by European funds under grants PGC2018-101836-B-I00 and ESP2017-87143-R (MINECO/FEDER), as well as by UK Space Agency through grant ST/R005761/1 and Italian Space Agency through grant 2018-2-HH.0. The IAA/CSIC team acknowledges financial support from the State Agency for Research of the Spanish MCIU through the 'Centre of Excellence Severo Ochoa' award for the Instituto de Astrofísica de Andalucía (SEV-2017-0709). This work was supported by the Belgian Fonds de la Recherche Scientifique – FNRS under grant numbers 30442502 (ET_HOME) and T.0171.16 (CRAMIC) and BELSPO BrainBe SCOOP Project.

## 12. Data Availability

All SPICE kernels used for the computations in this study are maintained by the ESA SPICE Service and available in the following repository: ftp://spiftp.esac.esa.int/data/SPICE in the dataset corresponding to each mission [Mars Express SPICE Kernel dataset] and [ExoMars2016 SPICE Kernel dataset]

Mars Express long term reference trajectory files, confirmed until January 2021, freely propagated until 2022:

*ORMF_T19_170101_220101_01400.BSP*, *ORMF_T19_181231_230101_01507.BSP*

Trace Gas Orbit long term reference trajectory files, confirmed until January 2021, freely propagated until 2024:

*em16_tgo_flp_001_02_20160314_20191102_v01.bsp, em16_tgo_flp_002_01_20180901_20240303_v01.bsp*

Solar System ephemeris and planetary constants files:

*mar097.bsp, de432s.bsp, pck00010.tpc*

## 13. References

Acton, C.H. et al: A look towards the future in the handling of space science mission geometry. Planetary and Space Science. Volume 150, Pages 9-12, 2018

Ao, C. O., C. D. Edwards Jr., D. S. Kahan, X. Pi, S. W. Asmar, and A. J. Mannucci, A first demonstration of Mars crosslink occultation measurements, Radio Sci. , 50, doi:10.1002/2015RS005750. 2015

Aoki, Shohei, et al. "Seasonal variation of the HDO/H2O ratio in the atmosphere of Mars at the middle of northern spring and beginning of northern summer." Icarus 260 (2015): 7-22. (https://doi.org/10.1016/j.icarus.2015.06.021)

Aoki S., Y.Sato, M. Giuranna, P. Wolkenberg, T.M. Sato, H. Nakagawa, Y. Kasaba, Mesospheric CO2 ice clouds on Mars observed by Planetary Fourier Spectrometer onboard Mars Express, 2018, vol.302, 175-190, Icarus, https://doi.org/10.1016/j.icarus.2017.10.047

Archinal, B. A. et al**,** Report of the IAU Working Group on Cartographic Coordinates and Rotational Elements: 2009, Celest Mech Dyn Astr, DOI 10.1007/s10569-010-9320-4, 2010

Ashman, M., Aberasturi, M., Cardesín, A., Costa, M., Frew, D., García, J., et al., "Science Planning Implementation and Challenges for the ExoMars Trace Gas Orbiter," SpaceOps 2018, 28 May - 1 June 2018, Marseille, France, AIAA, 2018.

Audouard, J., Poulet, F., Vincendon, M., Bibring, J. P., Forget, F., Langevin, Y., & Gondet, B. Mars surface thermal inertia and heterogeneities from OMEGA/MEX. Icarus, 233, 194-213. 2014

Barabash, S., Lundin, R., Andersson, H., Brinkfeldt, K., Grigoriev, A., Gunell, H., et al.: The Analyzer of Space Plasmas and Engergetic Atoms (ASPERA-3) for the Mars Express mission. Space Science Reviews, 126, 113–164. https://doi.org/10.1007/s11214-006-9124-8, 2006

Bertaux, J. L., Gondet, B., Lefèvre, F., Bibring, J. P., & Montmessin, F. (2012). First detection of O2 1.27 μm nightglow emission at Mars with OMEGA/MEX and comparison with general circulation model predictions. Journal of Geophysical Research: Planets, 117(E11).

Bibring, J.-P., et al., OMEGA: Observatoire pour la Minéralogie, l'Eau, les Glaces et l'Activité, ESA SP-1240, pp. 3–16. 2004

Billebaud, F., Brillet, J., Lellouch, E., Fouchet, T., Encrenaz, T., Cottini, V., ... & Forget, F.. Observations of CO in the atmosphere of Mars with PFS onboard Mars Express. Planetary and Space Science, 57(12), 1446-1457. 2009

Cardesin Moinelo, A. et al: ExoMars 2016 Trace Gas Orbiter and Mars Express Coordinated Science Operations Planning, Proceedings of SpaceOps Conference 2018

Carter, J., et al. Hydrous minerals on Mars as seen by the CRISM and OMEGA imaging spectrometers: Updated global view. Journal of Geophysical Research: Planets, 2013, vol. 118, no 4, p. 831-858. (doi:10.1029/2012JE004145)

Chicarro, A., Martin, P., & Trautner, R.: The Mars Express Mission: An Overview. ESA SP-1240, Mars Express: the scientific payload. ESA Publications Division, Noordwijk, Netherlands, ISBN 92-9092-556-6, p. 3 – 13, 2004

Costa, M., Cardesín A., Almeida M., Altobelli N., Constructing Optimized Observations – The Solar System Science Operations Laboratory, 63rd International Astronautical Congress, Naples, October 2012

Costa, M., "SPICE for ESA Planetary Missions: geometry and visualization support to studies, operations and data analysis within your reach", SpaceOps 2018, 28 May - 1 June 2018, Marseille, France, AIAA, 2018

Cox, C., Gérard, J.C., Hubert, B., Bertaux, J.L., Bougher, S.W., 2010. Mars ultraviolet dayglow variability: SPICAM observations and comparison with airglow model. J. Geophys. Res. 115, E04010. doi: 10.1029/20 09JE0 03504 .

Encrenaz, T., Fouchet, T., Melchiorri, R., Drossart, P., Gondet, B., Langevin, Y., ... & Bézard, B. (2006). Seasonal variations of the martian CO over Hellas as observed by OMEGA/Mars Express. Astronomy & Astrophysics, 459(1), 265-270.

Fedorova, A., Korablev, O., Bertaux, J.L., Rodin, A., Kiselev, A., Perrier, S.. Mars water vapor abundance from SPICAM IR spectrometer: seasonal and geographic distributions. J. Geophys. Res. 111, E09S08. doi: 10.1029/20 06JE0 02695 2006

Fedorova, A. A., Korablev, O. I., Bertaux, J. L., Rodin, A. V., Montmessin, F., Belyaev, D. A., & Reberac, A. (2009). Solar infrared occultation observations by SPICAM experiment on Mars-Express: Simultaneous measurements of the vertical distributions of H2O, CO2 and aerosol. Icarus, 200(1), 96-117.

Fedorova, A. A., Lefèvre, F., Guslyakova, S., Korablev, O., Bertaux, J. L., Montmessin, F., ... & Gondet, B. (2012). The O2 nightglow in the martian atmosphere by SPICAM onboard of Mars-Express. Icarus, 219(2), 596-608. doi:10.1016/j.icarus.2012.03.031

Fedorova, A. A., Montmessin, F., Rodin, A. V., Korablev, O. I., Määttänen, A., Maltagliati, L., & Bertaux, J. L. Evidence for a bimodal size distribution for the suspended aerosol particles on Mars. Icarus, 231, 239-260. 2014

Forget, F., Montmessin, F., Bertaux, J-L., Gonzalez-Galindo, F., Lebonnois, S., Quémerais, E., Réberac, A., Dimarellis, E., López-Valverde, M.A., 2009. Density and temperatures of the upper Martian atmosphere measured by stellar occul- tations with Mars Express SPICAM. J. Geophys. Res. 114, E01004. doi: 10.1029/ 20 08JE0 03086 .

Formisano, V. et al: The Planetary Fourier Spectrometer (PFS) onboard the European Mars Express mission, Planetary and Space Science, Volume 53, Issue 10, p. 963-974. 2005

Geiger, B et al: Long Term Planning for the ExoMars Trace Gas Orbiter Mission: Opportunity Analysis and Observation Scheduling, Proceedings of SpaceOps Conference, 2018

Giuranna M., S. Viscardy, F. Daerden, L. Neary, G. Etiope, D. Oehler, V. Formisano, A. Aronica, P. Wolkenberg, S. Aoki, A. Cardesin – Moinelo, J. Marin-Yaseli de la Parra, D. Merritt and M. Amoroso, Independent confirmation of methane spike on Mars and a source region east of Gale Crater, 2019, Nature Geoscience, 12, 326-332, https://doi.org/10.1038/s41561-019-0331-9 2019a

Giuranna M. P. Wolkenberg, D. Grassi, A. Aronica, S Aoki, D. Scaccabarozzi, B. Saggin, V. Formisano. The current weather and climate of Mars: 12 years of atmospheric monitoring by the Planetary Fourier Spectrometer on Mars Express, Icarus, accepted https://doi.org/10.1016/j.icarus.2019.113406 2019b

Grassi, D., Ignatiev, N.I., Zasova, L.V., Maturilli, A., Formisano, V., Bianchini, G.A., Giuranna, M., Methods for the analysis of data from the Planetary Fourier Spectrometer on the Mars Express mission. Planet. Space Sci. 53 (10), 1017–1034, 2005

Guerlet, S.; Ignatiev, N.; Fouchet, T.; Forget, F.; Millour, E.; Young, R.; Montabone, L.; Grigoriev, A.; Trokhimovskiy, A.; Montmessin, F.; Korablev, O. (2018), Thermal structure and aerosol content in the martian atmosphere from ACS-TIRVIM onboard ExoMars/TGO, European Planetary Science Congress 2018, held 16-21 September 2018 at TU Berlin, Berlin, Germany, id.EPSC2018-223.

Guerlet, S.; Young, R.; Ignatiev, N.; Forget, F.; Millour, E.; Fouchet, T.; Montabone, L.; Shakun, A.; Grigoriev, A.; Trokhimovskiy, A.; Montmessin, F.; Korablev, O. (2019), Investigation of the 2018 Global Dust Event from ACS-TIRVIM on board ExoMars/TGO, EPSC-DPS Joint Meeting 2019, held 15-20 September 2019 in Geneva, Switzerland, id. EPSC-DPS2019-776.

Ignatiev, N., Grigoriev, A., Shakun, A., Moshkin, B., Patsaev, D., Trokhimovskiy, A., ... & Guerlet, S.. Monitoring of the atmosphere of Mars with ACS TIRVIM nadir observations on ExoMars TGO. In European Planetary Science Congress (Vol. 12). 2018

Ignatiev, N. D. Grassi, S. Guerlet, P. Vlasov, A. Grigoriev, A. Shakun, A. Trokhimovskiy, O. Korablev, F. Montmessin, F. Forget, and L. Zasova (2019), Thermal structure and dust clouds during the 2018 dust storm from ACS-TIRVIM onboard ExoMars/TGO, Geophysical Research Abstracts Vol. 21, EGU2019-14988-1, 2019, EGU General Assembly 2019.

Ignatiev, N., A. Grigoriev, A. Shakun, B. Moshkin, D. Patsaev, A. Trokhimovskiy, O. Korablev, D. Grassi, P. Vlasov, L. Zasova, S. Guerlet, F. Forget, F. Montmessin, G. Arnold, O. Sazonov, A. Zharkov, I. Maslov, A. Kungurov, A. Santos-Skripko, and V. Shashkin and the ACS TIRVIM TEAM (2018), Monitoring of the atmosphere of Mars with ACS TIRVIM nadir observations on ExoMars TGO, European Planetary Science Congress 2018, 16–21 September 2018, Berlin, Germany, EPSC2018-891.

Jaumann, R. et al: The high-resolution stereo camera (HRSC) experiment on Mars Express: Instrument aspects and experiment conduct from interplanetary cruise through the nominal mission. Planetary and Space Science, Volume 55, Issues 7–8, Pages 928-952, May 2007

Korablev, O., Montmessin, F., Trokhimovskiy, A., Fedorova, A. A., Shakun, A. V., Grigoriev, A. V., et al., “The Atmospheric Chemistry Suite (ACS) of Three Spectrometers for the ExoMars 2016 Trace Gas Orbiter,” Space Science Reviews, Vol. 214, No. 7, https://doi.org/10.1007/s11214-017-0437-6, 2018

Korablev, O., Vandaele, A. C., Montmessin, F., Fedorova, A. A., Trokhimovskiy, A., ... & Erwin, J. T.. No detection of methane on Mars from early ExoMars Trace Gas Orbiter observations. Nature, 568(7753), 517, https://doi.org/10.1038/s41586-019-1096-4, 2019

G. Liuzzi, G. L. Villanueva, M. J. Mumma, M.l D. Smith, F. Daerden, B. Ristic, I. Thomas, A.C. Vandaele, M. R. Patel, J.-J. Lopez-Moreno, G. Bellucci, Methane on Mars: new insights into the sensitivity of CH4 with the NOMAD/ExoMars spectrometer through its first in-flight calibration, Icarus, 321, 671-690. 2019

López-Valverde, et al.: Investigations of the mars upper atmosphere with Exomars Trace Gas Orbiter. Space Science Reviews 214(1), 29. DOI 10.1007/s11214-017-0463-4. 2018

Luginin, M.; Fedorova, A.; Ignatiev, N.; Grigoriev, A.; Trokhimovskiy, A.; Shakun, A.; Montmessin, F.; Korablev, O. (2019), One year of observations of dust and water ice aerosols performed by ACS TIRVIM and NIR, EPSC-DPS Joint Meeting 2019, held 15-20 September 2019 in Geneva, Switzerland, id. EPSC-DPS2019-1316.

Määttänen, A., Montmessin, F., Gondet, B., Scholten, F., Hoffmann, H., González-Galindo, F., ... & Bibring, J. P.. Mapping the mesospheric CO2 clouds on Mars: MEx/OMEGA and MEx/HRSC observations and challenges for atmospheric models. Icarus, 209(2), 452-469. https://doi.org/10.1016/j.icarus.2010.05.017, 2010

Mahieux, A.C. Vandaele, S. Robert, V. Wilquet, R. Drummond, F. Montmessin, J.L. Bertaux, Densities and temperatures in the Venus mesosphere and lower thermosphere retrieved from SOIR on board Venus Express: Carbon dioxide measurements at the Venus terminator. J. Geophys. Res. 117, E07001 (2012)

Madeleine, J. B., Forget, F., Spiga, A., Wolff, M. J., Montmessin, F., Vincendon, M., ... & Schmitt, B. (2012). Aphelion water-ice cloud mapping and property retrieval using the OMEGA imaging spectrometer onboard Mars Express. Journal of Geophysical Research: Planets, 117(E11).

Mateshvili, N., Fussen, D., Vanhellemont, F., Bingen, C., Dodion, J., Montmessin, F., Perrier, S., Bertaux, J.L. Detection of Martian dust clouds by SPICAM UV nadir measurements during the October 2005 regional dust storm. Adv. Space Res. 40 (6), 869–880. doi: 10.1016/j.asr.2007.06.028 . 2007

Mateshvili, N., Fussen, D., Vanhellemont, F., Bingen, C., Dekemper, E., Loodts, N., & Tetard, C. Water ice clouds in the Martian atmosphere: Two Martian years of SPICAM nadir UV measurements. Planetary and Space Science, 57(8-9), 1022-1031. 2009

Melchiorri, R., Encrenaz, T., Fouchet, T., Drossart, P., Lellouch, E., Gondet, B., ... & Ignatiev, N.. Water vapor mapping on Mars using OMEGA/Mars Express. Planetary and Space Science, 55(3), 333-342. 2007

Metcalfe, L., et al.: ExoMars Trace Gas Orbiter (TGO) Science Ground Segment (SGS), Space Science Reviews, Volume 214, Issue 5, article id. 86, 26 pp., https://doi.org/10.1007/s11214-018-0522-5, 2018

Mitrofanov, I., Malakhov et al: Fine Resolution Epithermal Neutron Detector (FREND) Onboard the ExoMars Trace Gas Orbiter. Space Science Reviews, Volume 214, Issue 5, article id. 86, 26 pp., 2018

Montmessin, F., Gondet, B., Bibring, J. P., Langevin, Y., Drossart, P., Forget, F., & Fouchet, T.. Hyperspectral imaging of convective CO2 ice clouds in the equatorial mesosphere of Mars. Journal of Geophysical Research: Planets, 112(E11). 2007

Montmessin, F., Bertaux, J. L., Quémerais, E., Korablev, O., Rannou, P., Forget, F., ... & Dimarellis, E. (2006). Subvisible CO2 ice clouds detected in the mesosphere of Mars. Icarus, 183(2), 403-410.

Montmessin, F. et al, SPICAM on Mars Express: A 10 year in-depth survey of the Martian atmosphere. Icarus, Volume 297, p. 195-216. 2017

Montmessin F. and S. Ferron, (2018). A spectral synergy method to retrieve martian water vapor column-abundance and vertical distribution applied to Mars Express SPICAM and PFS nadir measurements, Icarus, 317, 549-569. https://doi.org/10.1016/j.icarus.2018.07.022

Ormston, T., Denis, M., Skuca, D., & Griebel, H. An ordinary camera in an extraordinary location: Outreach with the Mars webcam. Acta Astronautica, 69(7-8), 703–713. https://doi.org/10.1016/j.actaastro.2011.04.015, 2011

Pätzold, M. et al: Mars Express 10 years at Mars: Observations by the Mars Express Radio Science Experiment (MaRS). Planetary and Space Science, Volume 127, Pages 44-90. 2016

Picardi, G. et al. Radar soundings of the subsurface of Mars Science 310, 1925–1928, 2005

van der Plas, P., García-Gutiérrez, B., Nespoli, F., Pérez-Ayúcar, M., "MAPPS: A science planning tool supporting the ESA Solar System missions" SpaceOps 2016 Conference, AIAA, Daejeon, South Korea, 2016

Sánchez-Lavega, A., Chen-Chen, H., Ordonez-Etxeberria, I., Hueso, R., del Rio-Gaztelurrutia, T., Garro, A., ... & Wood, S. Limb clouds and dust on Mars from images obtained by the Visual Monitoring Camera (VMC) onboard Mars Express. Icarus, 299, 194-205. https://doi.org/10.1016/j.icarus.2017.07.026, 2018

Smith, M. D., Daerden, F., Neary, L., & Khayat, A. The climatology of carbon monoxide and water vapor on Mars as observed by CRISM and modeled by the GEM-Mars general circulation model. Icarus, 301, 117-131. 2018

Thomas, N., Cremonese, G., Ziethe, R., Gerber, M., Brändli, G., M Bruno, et al., "The Colour and Stereo Surface Imaging System (CaSSIS) for the ExoMars Trace Gas Orbiter," Space Science Reviews, Vol. 212, No. 3-4, pp. 1897–1944. 2018

Titov, D., Cardesin, A., Martin, P., Mars Express: Mission Status, Recent Findings and Future Plans, Proceedeings of the Sixth International Workshop on "Mars Atmosphere: Modelling and Observations", Granada, 17-20 January 2017

UPWARDS, Understanding Planet Mars, EU H2020 project: http://upwards.iaa.es

Vago, J., Witasse, O., Svedhem, H., Baglioni, P., Haldemann, A., Gianfiglio, G., et al., "ESA ExoMars program: The next step in exploring Mars," Solar System Research, Vol. 49, No. 7, pp. 518–528. 2015

Vandaele, A. C. et al: NOMAD, an Integrated Suite of Three Spectrometers for the ExoMars Trace Gas Mission: Technical Description, Science Objectives and Expected Performance, Space Science Reviews, 214: 80. https://doi.org/10.1007/s11214-018-0517-2, 2018

Vandaele, A. C., Korablev, O., et al: Martian dust storm impact on atmospheric H 2 O and D/H observed by ExoMars Trace Gas Orbiter. Nature, Volume 568, pages 521–525, 2019

Various authors: GRL Special Issue on the Effects of the Comet C/2013 A1 (Siding Spring) meteor shower in 2014 on Mars atmosphere and ionosphere: Observations from MAVEN, Mars Express, and Mars Reconnaissance Orbiter, Geophysical Research Letters, 2015, vol. 42, https://agupubs.onlinelibrary.wiley.com/doi/toc/10.1002/(ISSN)1944-8007.COMET1

Vincendon, M., Langevin, Y., Poulet, F., Bibring, J. P., Gondet, B., Jouglet, D., & OMEGA Team. (2008). Dust aerosols above the south polar cap of Mars as seen by OMEGA. Icarus, 196(2), 488-505.

Vincendon, M., Pilorget, Cedric; Gondet, Brigitte; Murchie, Scott; Bibring, Jean-Pierre, New near-IR observations of mesospheric CO2 and H2O clouds on Mars, JGR, Vol. 116 DOI: 10.1029/2011JE003827, 2011

Wolkenberg P., Marco Giuranna, Davide Grassi, Alessandro Aronica, Shohei Aoki, Diego Scaccabarozzi, Bortolino Saggin, Characterization of dust activity on Mars from MY27 to MY32 by PFS-MEX observations, vol. 310, 32-47, Icarus, doi:10.1016/j.icarus.2017.10.045. 2018

[Dataset] Mars Express SPICE kernel dataset. ftp://spiftp.esac.esa.int/data/SPICE/MARS-EXPRESS/ https://doi.org/10.5270/esa-esa-trn5vp1

[Dataset] ExoMars 2016 SPICE kernel dataset. ftp://spiftp.esac.esa.int/data/SPICE/ExoMars2016/ https://doi.org/10.5270/esa-pwviqkg